# Strong modulation of spin currents in bilayer graphene by static and fluctuating proximity exchange fields

Simranjeet Singh[1], Jyoti Katoch[1], Tiancong Zhu[1], Keng-Yuan Meng[1], Tianyu Liu[2], Jack T. Brangham[1], Fengyuan Yang[1], Michael Flatté[2] and Roland K. Kawakami[1]

[1]Department of Physics, The Ohio State University, Columbus, Ohio, United States, 43210

[2]Optical Science and Technology Center and Department of Physics and Astronomy, University of Iowa, Iowa City, Iowa, United States, 52242

**Abstract:**

Two dimensional (2D) materials provide a unique platform to explore the full potential of magnetic proximity driven phenomena, which can be further used for applications in next generation spintronic devices. Of particular interest is to understand and control spin currents in graphene by the magnetic exchange field of a nearby ferromagnetic material in graphene/ferromagnetic-insulator (FMI) heterostructures. Here, we present the experimental study showing the strong modulation of spin currents in graphene layers by controlling the direction of the exchange field due to FMI magnetization. Owing to clean interfaces, a strong magnetic exchange coupling leads to the experimental observation of complete spin modulation at low externally applied magnetic fields in short graphene channels. Additionally, we discover that the graphene spin current can be fully dephased by randomly fluctuating exchange fields. This is manifested as an unusually strong temperature dependence of the non-local spin signals in graphene, which is due to spin relaxation by thermally-induced transverse fluctuations of the FMI magnetization.

Use of the spin degree of freedom of electrons is poised to revolutionize next generation devices for logic [1] and memory [2] applications. Manipulation of spin current, using either a small electric or magnetic field, is the essential operation of such a device and is required to exploit the full versatility of spin related phenomena. Spins in graphene are of particular interest because of the fact that spins can propagate over large distances due to small spin-orbit (SO) coupling and negligible hyperfine interaction [3,4]. However, the absence of a strong SO field in graphene also means that spins in graphene cannot be manipulated by an external applied electric field [5]. In general, spins in graphene are manipulated by an out-of-plane magnetic field [6,7], known as Hanle spin precession, requiring large fields which are not viable for applications. An alternative route for efficient spin manipulation is to use the magnetic proximity effect of an adjacent ferromagnetic insulator (FMI). Two dimensional (2D) materials, like graphene, provide a unique platform to explore the proximity-induced phenomena as these effects are expected to be the strongest in 2D materials. There has been a great deal of interest to study the proximity effect induced changes in the electrical [8], optical [9,10] and spin [11] related properties of low dimensional materials. This research direction is further propelled by recent progress in the experimental techniques to assemble clean van der Waals heterostructures of 2D materials or mechanically transfer 2D samples onto arbitrary materials [12,13]. Recently, magnetic proximity effects in graphene/FMI heterostructures has been explored by charge transport measurements: (1) demonstration of ferromagnetism in graphene coupled to yttrium iron garnet (YIG), [14] and (2) large magnetic exchange fields experienced by charge carriers in graphene/EuS heterostructures [15]. Undoubtedly, these studies have established the presence of strong magnetic coupling across the interfaces of graphene and FMI materials, opening the doors for studying spin currents in graphene under the influence of magnetic proximity effect [16]. In particular, bilayer graphene is a system of choice for exploring these experiments due to the long spin diffusion lengths and spin lifetimes [17-19], electric field induced band gap engineering [20], and feasibility of electric field driven spin rotation [21].

In this Letter, we report the complete modulation of spin currents in bilayer graphene using the static and/or fluctuating components of the magnetic exchange field of an adjacent ferromagnet in a graphene/FMI heterostructure. For control of spin currents by a static exchange field, we employ a bilayer graphene lateral spin valve device on a YIG substrate and modulate the spin current in graphene by changing the direction of the YIG magnetization. A strong interfacial magnetic exchange coupling leads to the experimental observation of complete spin modulation in short graphene channels and at low magnetic fields. In addition, we discover that the spin current can be fully modulated by randomly fluctuating exchange fields. This is manifested as an unusually strong temperature dependence of the non-local spin signals, compared to the weak temperature dependence typically observed for graphene on non-magnetic substrates [3,22]. We attribute this to spin relaxation by thermally-induced transverse fluctuations of the YIG magnetization. These studies establish a lower bound on the magnetic exchange field to be ~1 Tesla.

We choose YIG for the ferromagnet because it is an insulator, has high Curie temperature, is chemically stable under ambient conditions, and is magnetically soft [23,24]. To prepare clean heterostructures of graphene/YIG we employ a dry transfer technique [13,25] as discussed in supplementary material [26]. The optical image of the hexagonal boron nitride (h-BN)/graphene stack on YIG is shown in Figure 1(a), where thin h-BN is highlighted by black dotted lines. The AFM topography of the h-BN/graphene/YIG surface is depicted in Figure 1(b) with the clean interface. In this structure, the h-BN serves as the tunnel barrier for spin injection into graphene [25]. Figure 1(c) shows the optical image of the device where the graphene and h-BN flakes are outlined with red and black dotted lines, respectively.

First, we establish the spin transport in a bilayer graphene channel (2.1 μm long and 2.2 μm, wide) on YIG by measuring the non-local magnetoresistance (MR). While sweeping an in-plane magnetic field, as schematically shown in inset to Figure 2(a), we record the non-local voltage signal ($V_{NL}$). Figure 2(a) shows $R_{NL}$, ($R_{NL} = V_{NL}/I$), as a function of in-plane magnetic field. The schematic of the experiment, to demonstrate control over spin currents in graphene by magnetic proximity effect, is shown in Figure 1(d). To modulate the spin signal in graphene, we align the magnetization of electrode E2 and E3 in either parallel (P) or antiparallel (AP) configuration and apply a fixed magnitude of magnetic field, $B_{ROT}$ = 15 mT, in the plane of the graphene. Note that this magnetic field is smaller than what is required to change or switch the electrode but large enough to saturate the YIG magnetization [19]. This magnetic field is rotated in the plane of the graphene by angle θ. Figure 2(b) shows $R_{NL}$ as a function of θ for the P configuration (blue circles) and AP configuration (red circles). We observe a clear modulation of the non-local signals for both P and AP configurations. To calculate the net change of the observed signal, we show in Figure 2(c) the differential $R_{NL}$ between the P and AP configurations. The change in spin signal due to the controlled change of YIG magnetization direction can be defined as: $\delta R = \frac{R(\theta=0°)-R(\theta)}{R(\theta=0°)}$ and one would expect to have maximum dephasing of the spins for $B_{ROT}$ applied at θ = 90°. This is indeed what we observe as the non-local MR signal goes to zero for magnetic field applied at θ = 90° and corresponds to 100% modulation. In other words, when the YIG magnetization is transverse to the injected spin polarization, there is a complete dephasing of the injected spins in graphene channel.

By performing control experiments, we show that this modulation is primarily due to the proximity exchange field, $B_{ex}$ (~ $M_{YIG}$), originating from quantum mechanical interactions of the carriers in graphene with the YIG magnetization, as opposed to a direct effect of the external field $B_{ROT}$. One possible effect of $B_{ROT}$ is to tilt the Co magnetizations asymmetrically to reduce $R_{NL}$. This effect is ruled out through anisotropic magnetoresistance measurements of the Co electrodes [34,35], as discussed in the supplementary material [26]. The other possible effect of $B_{ROT}$ is the direct interaction with the carriers in graphene to dephase the spin polarization via Hanle spin precession. Indeed, with the presence of proximity exchange field, the Hanle spin precession should be governed by the total magnetic field $B_{total} = B_{ex} + B_{ROT}$. The relative importance of $B_{ex}$ and $B_{ROT}$ can be determined by performing angular scans (θ) for different magnitudes of $B_{ROT}$. If $B_{ex}$ dominates, there should be very little dependence on $|B_{ROT}|$ because $M_{YIG}$ is fully saturated for fields higher than few mT [26] and $B_{ex}$ is proportional to $M_{YIG}$. If the direct interaction of $B_{ROT}$ dominates, then the modulation should become stronger with increasing $|B_{ROT}|$. Figure 3(a) shows the angular scan of $R_{NL}$ vs. θ for different values of $|B_{ROT}|$ from 6 to 18 mT. The most striking feature is the similarity of all the curves, which show full modulation even for the lower applied magnetic fields. This indicates that the modulation is dominated by the proximity exchange field. We further test this conclusion by performing the same measurement of a control sample consisting of a bilayer graphene spin valve on $SiO_2/Si(001)$ substrate. The measured non-local MR signal is shown in Figure 3(b). The non-local signal as a function of θ for different values of $|B_{ROT}|$, measured for both parallel and anti-parallel configurations of injector/detector electrodes, is shown in Figure 3(c). Clearly, we observe a highest modulation of only a few percent (~ 10%) in contrast to the 100% modulation when graphene is placed on a YIG substrate. Furthermore, we have also carried out spin modulation experiment on an another control sample, wherein graphene is separated from YIG by a thin h-BN (gra/h-BN/YIG) and we do not observe nonlocal spin signal modulation more than a few percent (supplementary file) [26].Thus, for graphene on a non-magnetic substrate, the modulation by $B_{ROT}$ is much weaker and has a strong dependence on the magnitude of the field consistent with Hanle effect.

Next, we study the temperature dependence of spin signal in the graphene channel coupled to YIG, which reveals a new mechanism for spin relaxation due to fluctuating proximity exchange fields. The magnitude of the measured MR signal ($\Delta R_{NL}$) is defined as the difference of $R_{NL}$ between the parallel and antiparallel configurations (Figure 2a), and the measured value is approximately 0.22 Ω at 15 K. Then, we measure $\Delta R_{NL}$ at different temperatures and the observed data is shown in Figure 4(a). The spin signal in graphene on a non-magnetic substrates normally has a weak temperature dependence and decreases approximately by a factor of 2 (or so) going from 10 K to room temperature [22,36,37]. However, as clearly seen from Figure 4(a), we observe that the spin signal rapidly decays as temperature increases, and completely disappears at ~230 K. Because the non-local spin signal is known to be dependent on the graphene resistivity ρ and the interfacial contact resistances of the electrodes [3], we first check whether these can account for the observed temperature dependence of $\Delta R_{NL}$. The temperature dependence of the graphene sheet resistance (or resistivity) on YIG is shown in Figure 4(b) and is similar to what has been widely reported for graphene on other non-magnetic substrates [12,22,38,39]. We also point out that the interfacial contact resistances of both injector and detector electrodes stay constant over the measured temperature range as shown in Figure 4(c). This rules out that the strong temperature dependent decay of the spin signal is merely due to changes in ρ or the contact resistances. Additionally, we have measured the temperature dependence of MR signals in the gra/h-BN/YIG control sample (supplementary file) and did not observe a strong temperature dependence [26]. In the following we argue that the observed temperature dependence of the MR spin signal in graphene/YIG can be explained by the electron spin dephasing in graphene due to the random transverse magnetization fluctuations of the YIG film. To qualitatively understand this unusual temperature dependence of the spin signal, we consider the interaction between conduction electrons in graphene and magnetization of YIG. The terms in the Hamiltonian associated with the conduction electron spins are given by:

$$H_e = A_{ex} \cdot g_e \mu_B \vec{S}_e \cdot \langle \vec{M} \rangle + g_e \mu_B \vec{S}_e \cdot \vec{B}_{app} = g_e \mu_B \vec{S}_e \cdot (\langle \vec{B}_{ex} \rangle + \vec{B}_{app}) = g_e \mu_B \vec{S}_e \cdot \langle \vec{B}_{eff} \rangle, \quad (1)$$

where $A_{ex}$ is the proximity induced exchange coupling strength between YIG and graphene, $\vec{M}$ is the YIG magnetization, and $\langle \vec{B}_{ex} \rangle = A_{ex} \langle \vec{M} \rangle$ is the effective exchange field. The averaging $\langle ... \rangle$ is over the ensemble of magnetic moments in YIG that are in proximity with graphene. At a finite temperature, $\vec{M}$ in YIG fluctuates, which in turn causes the proximity exchange field in graphene to fluctuate as well. For an electron travelling through graphene, the time and spatial variation of magnetization in YIG results in a varying effective magnetic field acting on the electron spin. This varying effective magnetic field can be modeled as a time-dependent, randomly fluctuating magnetic field $\vec{B}_{ex}(t) = \langle \vec{B}_{ex} \rangle + \Delta \vec{B}_{ex}$. Previous theoretical work had predicted that the randomly fluctuating magnetic field can cause extra spin relaxation [40,41] and has been used to explain spin transport phenomena in graphene decorated with paramagnetic hydrogen adatoms [42]. Furthermore, the fluctuation strength of YIG magnetization is expected to be temperature dependent. As a result, the spin relaxation rate caused by the magnetization fluctuation should be temperature dependent as well. In the following, we use the above model to understand the observed temperature dependence data. For non-local geometry [Figure 1(d)], the injected spin polarization, the applied magnetic field, and the effective exchange field lie along the same axis (*y* axis in our case). The spin relaxation rate induced by the random fluctuating field is given by the longitudinal spin relaxation term:

$$\frac{1}{\tau_1^{ex}} = \frac{(\Delta B_{tr})^2}{\tau_c} \frac{1}{(B_{app,y}+\bar{B}_{ex,y})^2+(\gamma_e\tau_c)^{-2}} \approx \frac{(\Delta B_{tr})^2}{\tau_c} \frac{1}{(\bar{B}_{ex,y})^2+(\gamma_e\tau_c)^{-2}}, \qquad (2)$$

where $(\Delta B_{tr})^2 = (\Delta B_{ex,x})^2 + (\Delta B_{ex,z})^2$ is the fluctuation of exchange field in the transverse direction, $B_{app,y}$ is ignored as $\bar{B}_{ex,y} \gg B_{app,y}$, $\gamma_e$ is the gyromagnetic ratio of electron, and $\tau_c$ is the correlation time of the exchange field fluctuation defined as: $\langle \Delta \vec{B}_{ex}(t) \cdot \Delta \vec{B}_{ex}(t-t') \rangle_t \propto exp\,(-t/\tau_c)$. The exchange field fluctuation in graphene should be strongly associated with the magnetization fluctuation of YIG. At finite temperature, thermally driven magnetization fluctuations suppress the equilibrium magnetization from the saturated 0 K value. Assuming that transverse magnetization fluctuations in YIG are responsible for the reduction of M with increasing temperature, we can rewrite Eq. (2) as:

$$\frac{1}{\tau_1^{ex}} = \frac{A_{ex}^2\big((M_0)^2-(\bar{M}_y(T))^2\big)}{\tau_c\big[(A_{ex}\bar{M}_y(T))^2+(\gamma_e\tau_c(T))^{-2}\big]}, \qquad (3)$$

where $M_0$ is the saturation magnetization of YIG at 0 K, $\bar{M}_y(T)$ is the temperature dependent equilibrium magnetization in the $y$ direction, and $\tau_c(T)$ is the temperature dependent correlation time. We extract the temperature dependence of $\bar{M}_y$ from the measured temperature dependence of saturation magnetization of YIG up to 300 K [Figure 4(d)]. Previous study of bulk YIG shows that the reduction of saturation magnetization follows $\sim T^{3/2}$ in the low temperature regime (<25 K), while it follows a $\sim T^3$ in the higher temperature regime (25 K~250 K) [43]. We fit the measured data with both terms, and find that the contribution of the $T^3$ term is minimal. To simplify the spin transport equation later, we assume that:

$$\frac{\bar{M}_y}{M_0} = 1 - aT^{\frac{3}{2}}, \qquad (4)$$

and get $a = 6.314\times10^{-5}$ K$^{-3/2}$ from fitting with experimental YIG magnetization.

To obtain the temperature dependence of correlation time, we have adapted a macroscopic picture of local magnetization fluctuation which has been developed through fluctuation-dissipation theorem and had successfully explained spin Seebeck effect in Pt/YIG structure [44-46]. As explained in detail in the supplementary file [26], the relationship between correlation time and YIG magnetization is:

$$\frac{1}{\tau_c(T)} = \frac{\alpha}{\sqrt{1+\alpha^2}}\omega_0 = \frac{\alpha\gamma H_0}{\sqrt{1+\alpha^2}} = \eta M_{YIG}(T). \tag{5}$$

To simplify the expression of $\tau_1^{ex}$, we put Eqs. (4) and (5) into Eq. (3):

$$\frac{1}{\tau_1^{ex}} = \frac{A_{ex}^2\left((M_0)^2 - (\bar{M}_y(T))^2\right)}{\tau_c\left[(A_{ex}\bar{M}_y(T))^2 + (\gamma_e\tau_c(T))^{-2}\right]} = \frac{1-(\bar{M}_y/M_0)^2}{\bar{M}_y/M_0} \cdot \frac{\eta(\gamma_e A_{ex})}{(\gamma_e A_{ex})^2+\eta^2} \cdot \gamma_e A_{ex} M_0. \tag{6}$$

We define $\xi(T) = \frac{1-(\bar{M}_y/M_0)^2}{\bar{M}_y/M_0}$ which is the only temperature dependent term, and rewrite the whole equation as:

$$\frac{1}{\tau_1^{ex}} = \xi(T) \cdot \frac{\eta(\gamma_e A_{ex})}{(\gamma_e A_{ex})^2+\eta^2} \cdot \gamma_e A_{ex} M_0 \tag{7}$$

The non-local spin signal measured in the graphene lateral spin valve device can be written as [32,37]:

$$R_{NL} = p_1 p_2 R_N e^{-L/\lambda}, \tag{8}$$

where $p_1, p_2$ are the spin polarizations at the Co/h-BN/graphene injector and detector junctions, respectively, $\lambda = \sqrt{D\tau_{total}}$ is the spin diffusion length, $R_N$ is the spin resistance of the graphene channel, $D$ is the diffusion constant, and $\tau_{total}$ is the spin lifetime of electron spins in graphene. Apart from the spin relaxation mechanism in graphene on a non-magnetic substrate, in our case we have additional spin relaxation, $\frac{1}{\tau_1^{ex}}$, caused by the YIG magnetization fluctuations [Eq. (3)]. Thus, Eq. (8) becomes:

$$R_{NL} = p_1 p_2 R_N e^{-L\cdot\left(\frac{1}{D\tau_1^{ex}} + \frac{1}{\lambda_{int}^2}\right)^{-1/2}}, \tag{9}$$

where $\lambda_{int}$ is the spin diffusion length of graphene for the case of a non-magnetic substrate. Using Eq. (9), we obtain:

$$R_{NL} = \mathcal{R}e^{-L\cdot\left(\frac{\xi(T)}{\beta} + \frac{1}{\lambda_{int}^2}\right)^{-1/2}}, \tag{10}$$

where $\frac{1}{\beta} = \frac{\gamma_e}{D} \frac{\eta(\gamma_e A_{ex})}{(\gamma_e A_{ex})^2 + \eta^2} \cdot A_{ex} M_0$. Using L = 2.1 $\mu m$ (channel length), we fit the observed temperature dependence of non-local MR signal. The model fits very well with the experimental data as shown in Figure 4(a), from which we can extract $\mathcal{R} = 0.7015 \, \Omega, \lambda_{int} = 1.9561 \, \mu m,$ and $\beta = 1.5578 \times 10^{-13} m^2$.

To calculate the exchange field in graphene at 0 K, we focus on the $\beta$ coefficient from the fitting using:

$$B_{ex}(0) = A_{ex} M_0 = \sqrt{\frac{D}{\eta \beta M_0 - D}} \cdot \frac{\eta M_0}{\gamma_e} = \frac{1}{\gamma_e} \cdot \sqrt{\frac{1}{(\frac{\beta}{D})\tau_c - \tau_c^2}}, \tag{11}$$

where $\tau_c$ is the correlation time at 0 K. Assuming a typical $D = 0.015 \, m^2/s$ for graphene [3,25,37], we plot $B_{ex}(0)$ as function of different $\tau_c$ as shown in Figure 4(e). Our model gives a lower bound of 1 Tesla of the exchange field. We also measure the temperature dependence of the spin signal modulation and observe a clear 100 % signal modulation up to ~150 K where we have clear MR signals, confirming the existence of this magnetic proximity induced phenomena at higher temperatures [26].

In conclusion, we have experimentally demonstrated the full modulation of spins in graphene by employing the proximity exchange fields present at the interface of graphene/FMI heterostructure. The observed strong temperature dependence of non-local MR signals in graphene spin valves for the first time experimentally establishes the additional spin dephasing mechanism due to the magnetic fluctuations in graphene/ferromagnet systems. We have used this novel observation to extract a lower bound of the interfacial magnetic exchange field. The work presented here will further help understand (and also exploit) the interfacial effects due to interaction of spins and magnetization in ferromagent/non-magnetic bilayer systems in general.

**Acknowledgements:**

S.S., J.K., T.Z., and R.K.K. acknowledge support from ONR (No. N00014-14-1-0350), NSF (No. DMR-1310661), ARO (W911NF-11-1-0182) and C-SPIN, one of the six SRC STARnet Centers, sponsored by MARCO and DARPA. K.Y.M. acknowledges support from NSF (No. DMR-1507274). T.L. and M.F. acknowledge support from C-SPIN, one of the six SRC STARnet Centers, sponsored by MARCO and DARPA. J.T.B. and F.Y.Y. acknowledge support from DOE, Office of Science, Basic Energy Sciences, under Award No. DE-SC0001304. We also thank Walid Amamou, Jinsong Xu and Igor Pinchuk for technical assistance.**References:**

[1] B. Behin-Aein, D. Datta, S. Salahuddin, and S. Datta, Nature Nanotechnology **5**, 266 (2010).
[2] L. Liu, C. F. Pai, Y. Li, H. W. Tseng, D. C. Ralph, and R. A. Buhrman, Science **336**, 555 (2012).
[3] W. Han, R. K. Kawakami, M. Gmitra, and J. Fabian, Nature Nanotechnology **9**, 794 (2014).
[4] M. Wojtaszek, I. J. Vera-Marun, E. Whiteway, M. Hilke, and B. J. van Wees, Physical Review B **89**, 035417 (2014).
[5] H. C. Koo, J. H. Kwon, J. Eom, J. Chang, S. H. Han, and M. Johnson, Science **325**, 1515 (2009).
[6] N. Tombros, C. Jozsa, M. Popinciuc, H. T. Jonkman, and B. J. van Wees, Nature **448**, 571 (2007).
[7] W. Han, K. Pi, W. Bao, K. M. McCreary, Y. Li, W. H. Wang, C. N. Lau, and R. K. Kawakami, Applied Physics Letters **94**, 222109 (2009).
[8] C. R. Dean, L. Wang, P. Maher, C. Forsythe, F. Ghahari, Y. Gao, J. Katoch, M. Ishigami, P. Moon, M. Koshino, T. Taniguchi, K. Watanabe, K. L. Shepard, J. Hone, and P. Kim, Nature **497**, 598 (2013).
[9] S. Wu, L. Wang, Y. Lai, W.-Y. Shan, G. Aivazian, X. Zhang, T. Taniguchi, K. Watanabe, D. Xiao, C. Dean, J. Hone, Z. Li, and X. Xu, Science Advances **2**, 5 (2016).
[10] P. Rivera, K. L. Seyler, H. Yu, J. R. Schaibley, J. Yan, D. G. Mandrus, W. Yao, and X. Xu, Science **351**, 688 (2016).
[11] M. H. D. Guimarães, P. J. Zomer, J. Ingla-Aynés, J. C. Brant, N. Tombros, and B. J. van Wees, Physical Review Letters **113**, 086602 (2014).
[12] C. R. Dean, A. F. Young, I. Meric, C. Lee, L. Wang, S. Sorgenfrei, K. Watanabe, T. Taniguchi, P. Kim, K. L. Shepard, and J. Hone, Nature Nanotechnology **5**, 722 (2010).
[13] L. Wang, I. Meric, P. Y. Huang, Q. Gao, Y. Gao, H. Tran, T. Taniguchi, K. Watanabe, L. M. Campos, D. A. Muller, J. Guo, P. Kim, J. Hone, K. L. Shepard, and C. R. Dean, Science **342**, 614 (2013).
[14] Z. Wang, C. Tang, R. Sachs, Y. Barlas, and J. Shi, Physical Review Letters **114**, 016603 (2015).
[15] P. Wei, S. Lee, F. Lemaitre, L. Pinel, D. Cutaia, W. Cha, F. Katmis, Y. Zhu, D. Heiman, J. Hone, J. S. Moodera, and C. T. Chen, Nature Materials **15**, 711 (2016).
[16] L. Johannes Christian, A. K. Alexey, W. Magdalena, and J. v. W. Bart, 2D Materials **4**, 014001 (2017).
[17] T. Y. Yang, J. Balakrishnan, F. Volmer, A. Avsar, M. Jaiswal, J. Samm, S. R. Ali, A. Pachoud, M. Zeng, M. Popinciuc, G. Güntherodt, B. Beschoten, and B. Özyilmaz, Physical Review Letters **107**, 047206 (2011).
[18] W. Han and R. K. Kawakami, Physical Review Letters **107**, 047207 (2011).
[19] A. Avsar, I. J. Vera-Marun, J. Y. Tan, G. K. W. Koon, K. Watanabe, T. Taniguchi, S. Adam, and B. Ozyilmaz, NPG Asia Materials **8**, e274 (2016).
[20] Y. Zhang, T. T. Tang, C. Girit, Z. Hao, M. C. Martin, A. Zettl, M. F. Crommie, Y. R. Shen, and F. Wang, Nature **459**, 820 (2009).
[21] P. Michetti, P. Recher, and G. Iannaccone, Nano Letters **10**, 4463 (2010).
[22] M. V. Kamalakar, A. Dankert, J. Bergsten, T. Ive, and S. P. Dash, Scientific Reports **4**, 6146 (2014).
[23] H. L. Wang, C. H. Du, Y. Pu, R. Adur, P. C. Hammel, and F. Y. Yang, Physical Review B **88**, 100406 (2013).
[24] H. L. Wang, C. H. Du, Y. Pu, R. Adur, P. C. Hammel, and F. Y. Yang, Physical Review Letters **112**, 197201 (2014).
[25] S. Singh, J. Katoch, J. Xu, C. Tan, T. Zhu, W. Amamou, J. Hone, and R. Kawakami, Applied Physics Letters **109**, 122411 (2016).10

**Figure Captions:**

**Figure 1**: (a) Optical image of an h-BN/graphene stack on a YIG substrate (b) Atomic force microscopy image of the h-BN/graphene/YIG heterostructure surface after vacuum annealing, showing the clean surface. (c) Optical image of the completed spin valve device. The red and black dotted lines in (a), (b) and (c) outlines the graphene and h-BN tunnel barrier boundaries, respectively. (d) Schematic of the experiment used to demonstrate spin current modulation in graphene. A magnetic field ($B_{ROT}$) applied at different θ defines the YIG magnetization ($M_{YIG}$) relative to the magnetization of Co injector/detector electrodes (or injected spin polarization in graphene).

**Figure 2**: Spin signal modulation in graphene coupled to a YIG substrate at 15 K. (a) The measured non-local MR signal in a graphene spin valve on YIG. The blue and red arrows represent the relative magnetization direction of injector (E2) and detector (E3) electrodes. Inset: schematic of the non-local spin valve measurement setup. (b) Non-local MR signal measured as function of $B_{ROT}$ magnetic field direction (θ). A fixed $B_{ROT}$ = 15 mT is applied in the YIG plane. The blue and red filled circles show the measured data for parallel and anti-parallel configuration of the injector/detector electrodes, respectively. (c) Differential non-local MR between the parallel and anti-parallel data from (b) as a function of θ, showing that for θ = 90°, the signal goes to zero which indicates a complete spin dephasing.

**Figure 3**: Dependence of spin signal modulation on the magnitude of $B_{ROT}$. (a) Spin signal modulation, $R_{NL}$[P-AP], for $B_{ROT}$ ranging from 6 to 18 mT for a graphene device on YIG shows that the spin signal modulation is independent of the magnitude of $B_{ROT}$. (b) Non-local MR signal for a bilayer graphene on a non-magnetic $SiO_2$/Si substrate with the relative magnetization orientations of the electrodes denoted by the red and blue arrows. (c) Spin signal modulation as function of θ for a graphene device on $SiO_2$/Si at different applied $B_{ROT}$ fields between 6 and 15 mT for both parallel and anti-parallel configurations.

**Figure 4**: (a) Temperature dependence of non-local MR signal in a graphene spin valve on YIG, where the red filled squares are experimental data and the blue solid line is the fitting by a model based on spin dephasing due to the temperature dependent transverse magnetization fluctuations of YIG. (b) Temperature dependence of graphene sheet resistance is. (c) Temperature dependence of interfacial contact resistances of the injector (black) and detector (red) electrodes. (d) Temperature dependence of saturation magnetization (red filled circles) of the YIG film extracted from magnetization measurements. The solid blue line is a fitting of the temperature dependent magnetization data by Eq. (4). (e) Extracted exchange field as function of correlation time of fluctuating YIG magnetization.

Figure 1

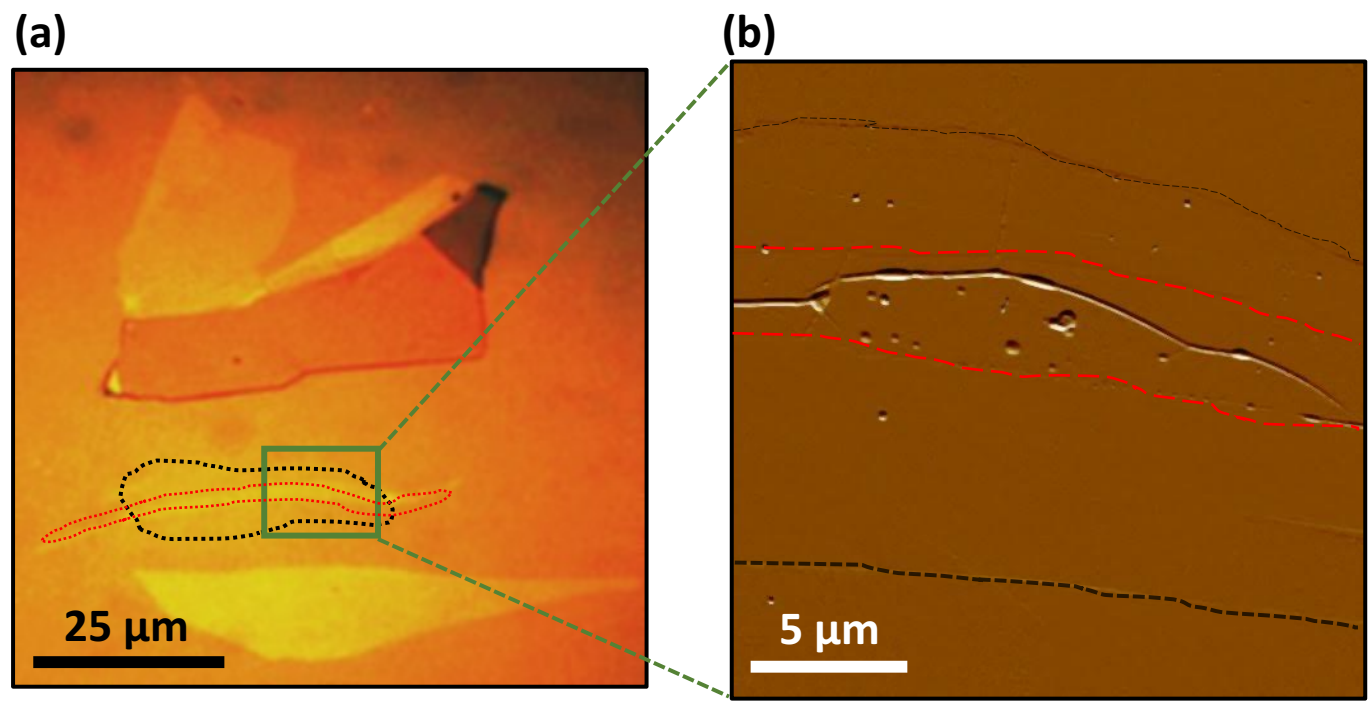
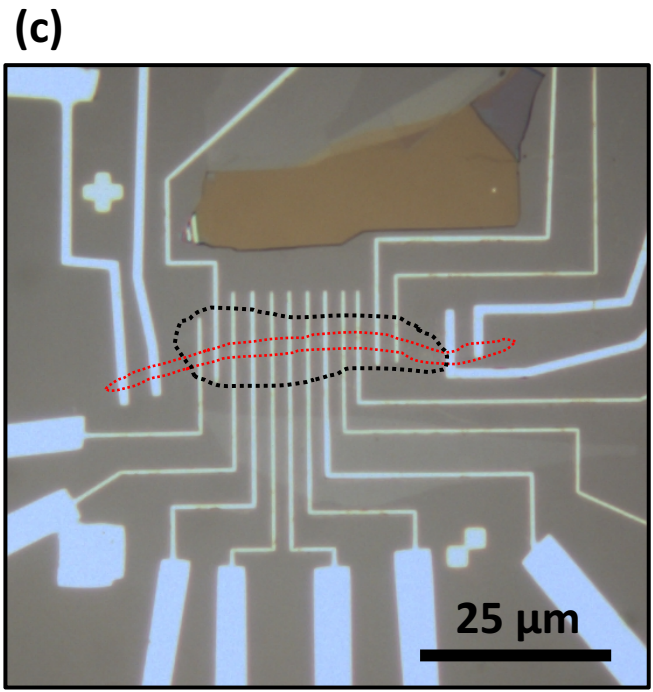
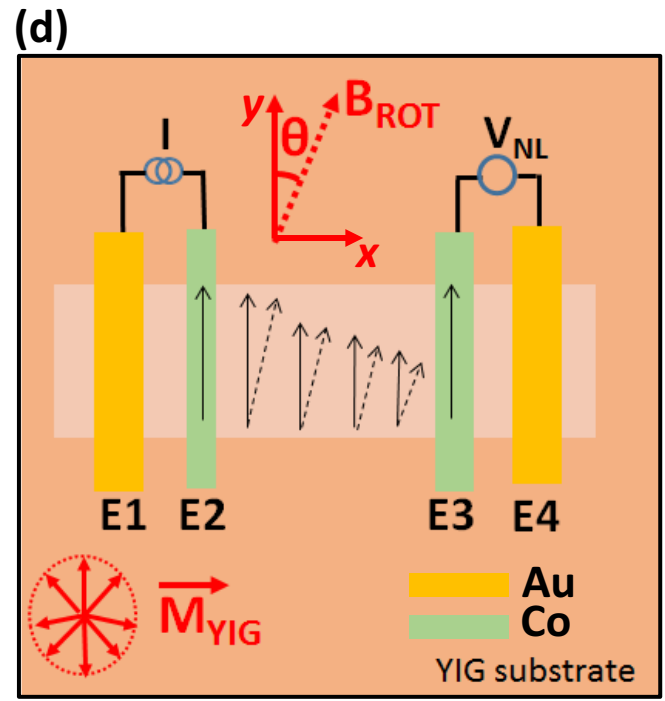

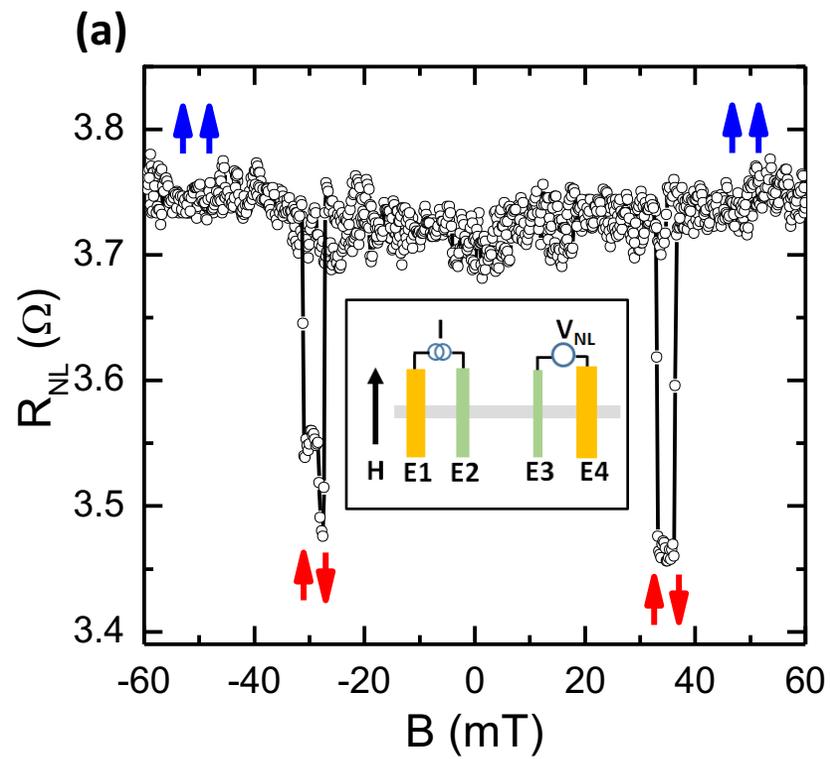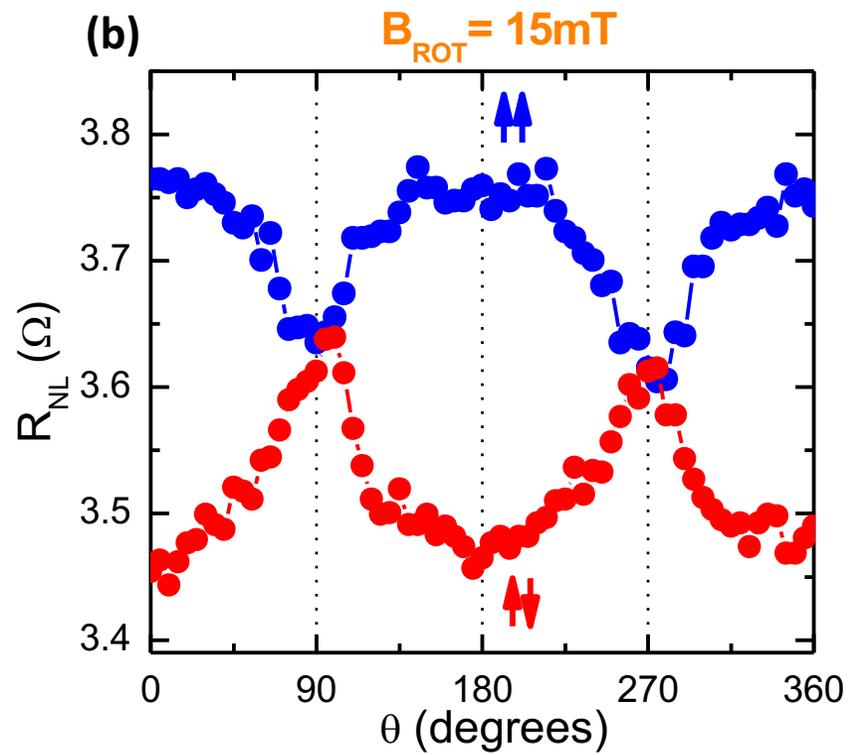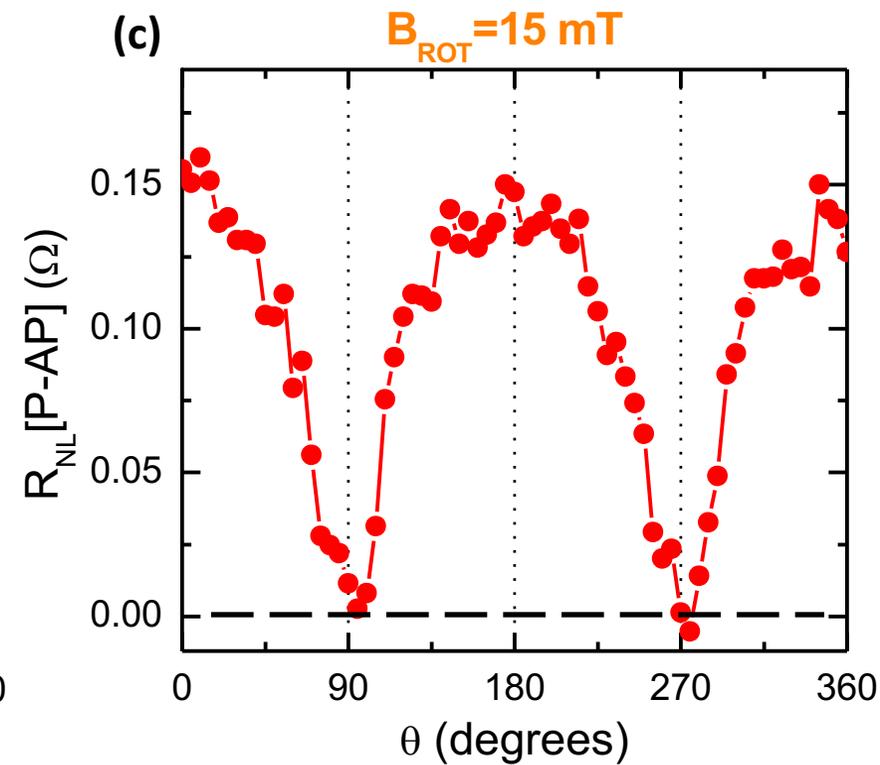

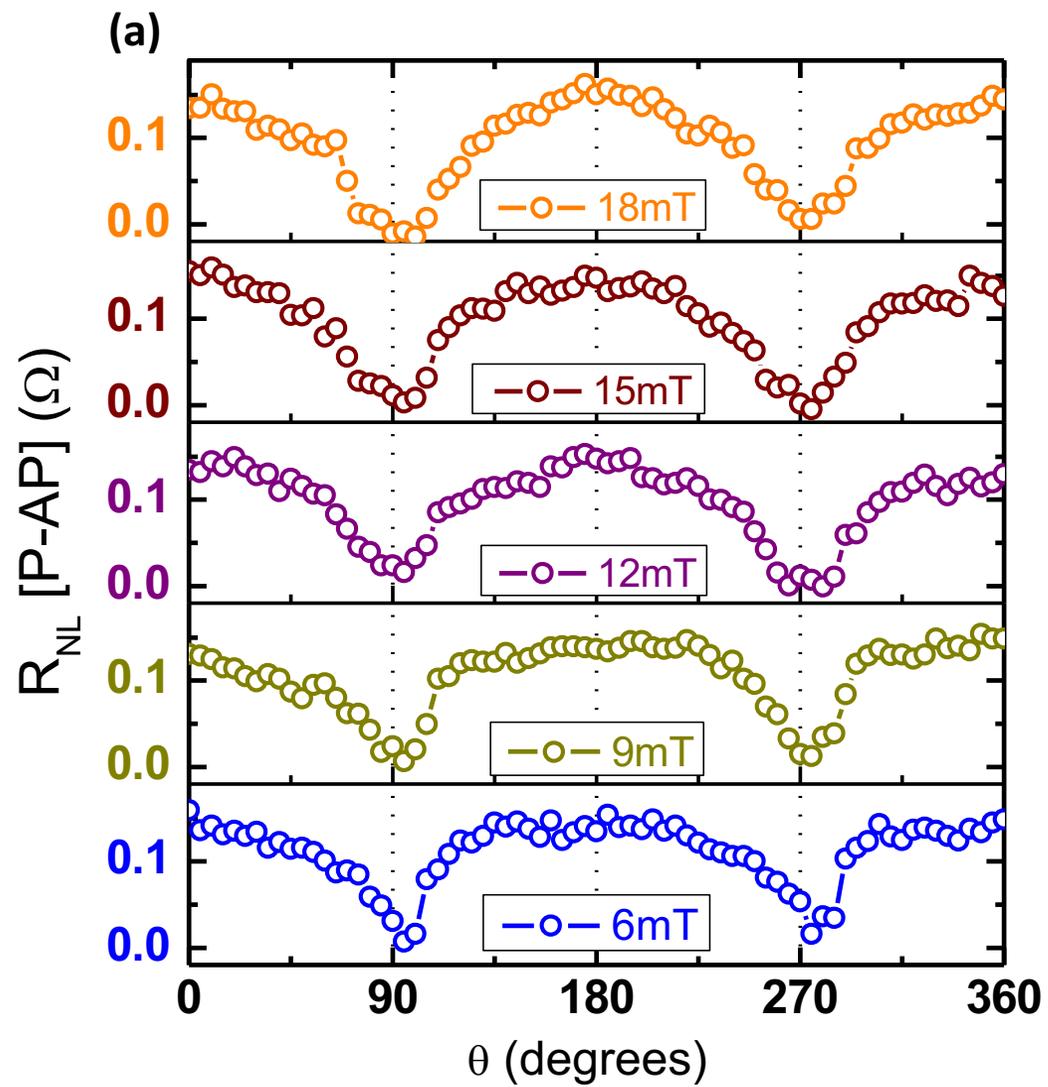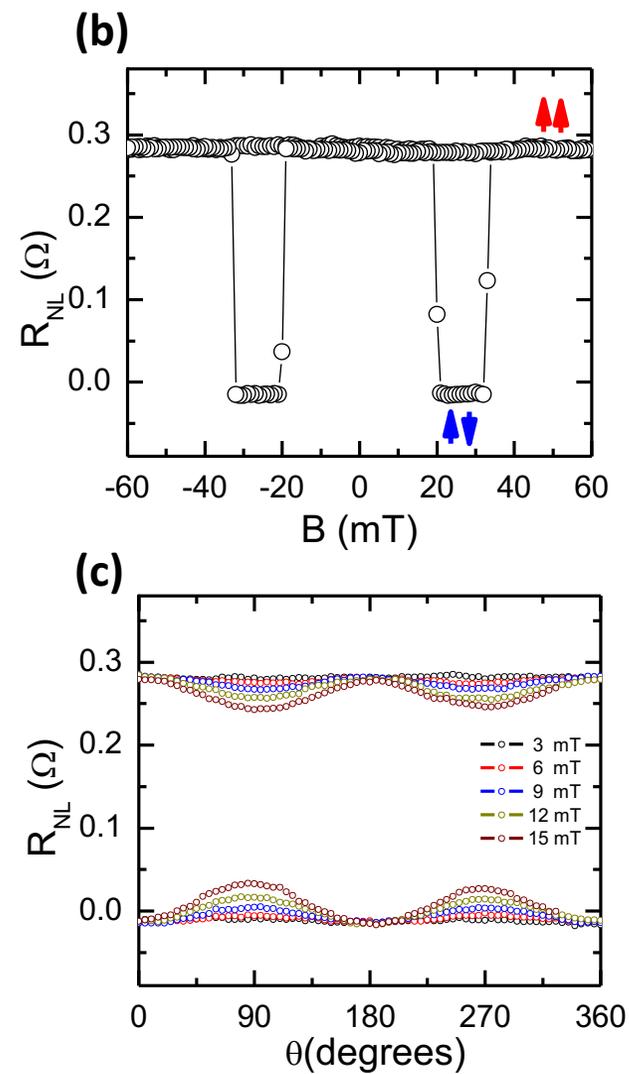

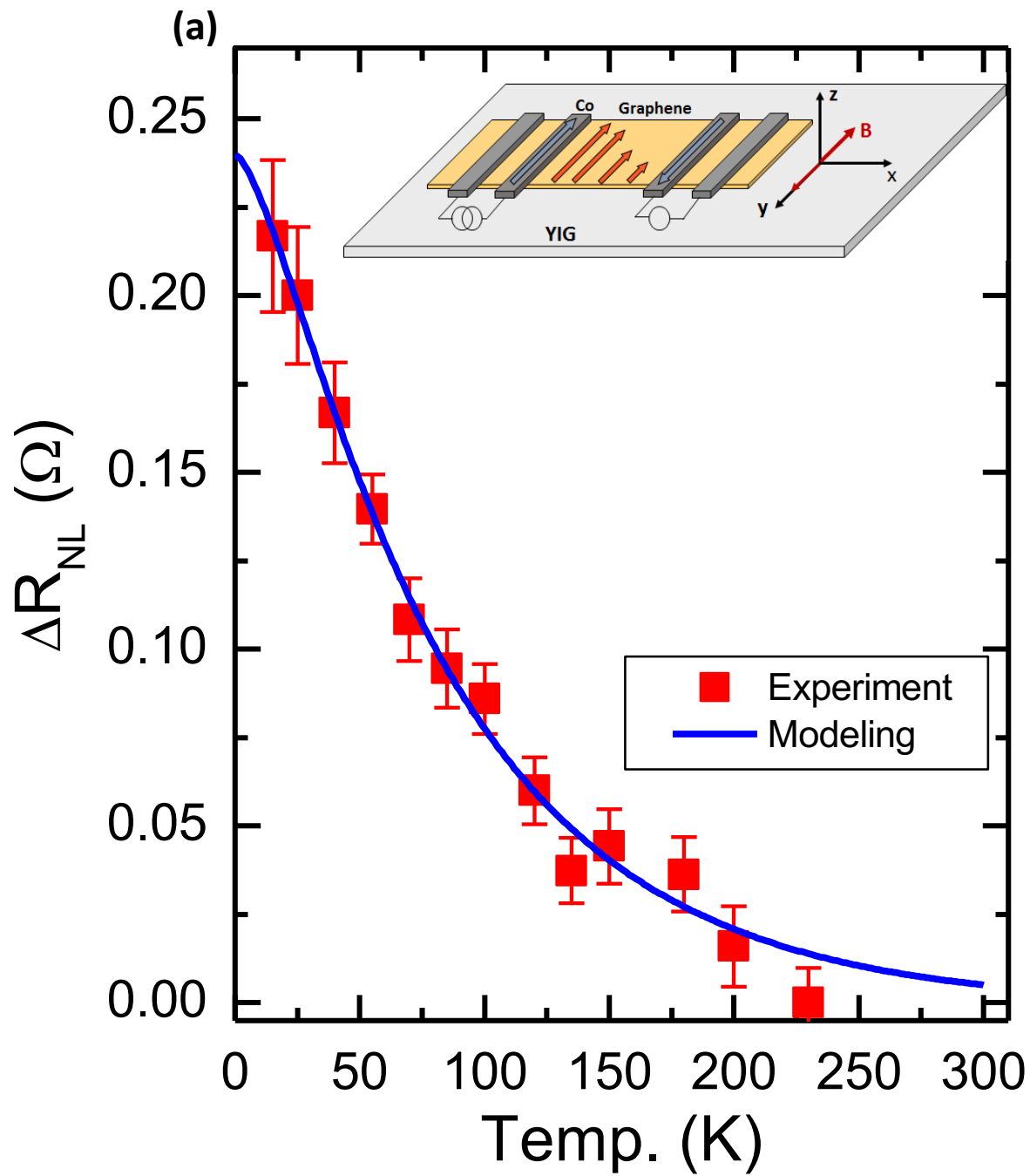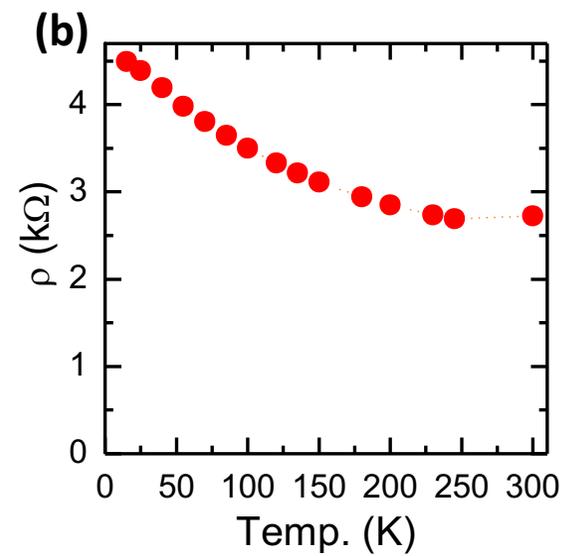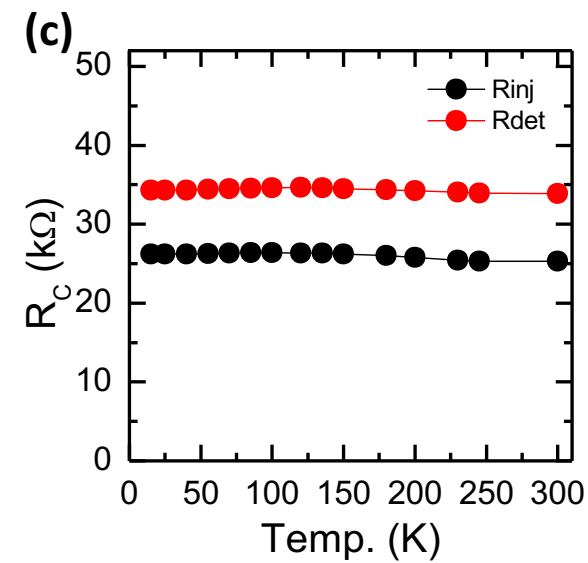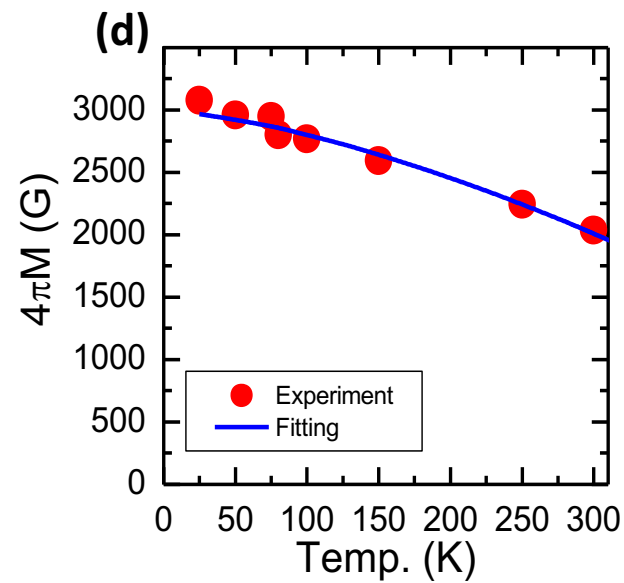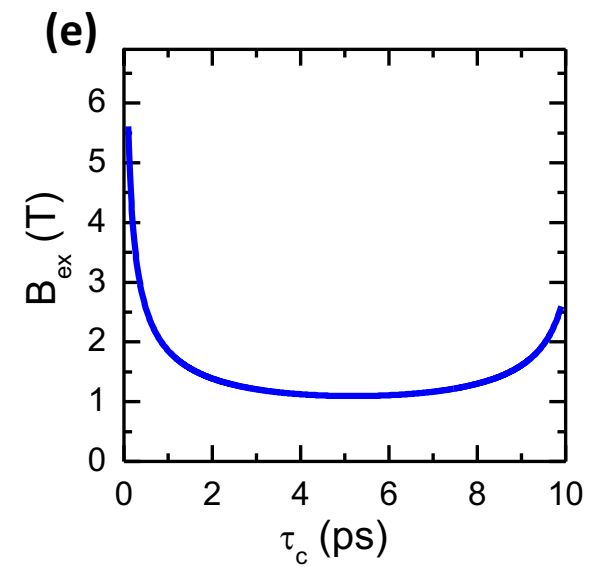

# Supplementary information

# Strong modulation of spin currents in bilayer graphene by static and fluctuating proximity exchange fields

Simranjeet Singh[1], Jyoti Katoch[1], Tiancong Zhu[1], Keng-Yuan Meng[1], Tianyu Liu[2], Jack T. Brangham[1], Fengyuan Yang[1], Michael Flatté[2] and Roland K. Kawakami[1]

[1]*Department of Physics, The Ohio State University, Columbus, Ohio, United States, 43210*

[2]*Optical Science and Technology Center and Department of Physics and Astronomy, University of Iowa, Iowa City, Iowa, United States, 52242*

1. **Heterostructure preparation and magnetization hysteresis loop of YIG**

High quality 20-nm thick YIG epitaxial films are grown on (111)-oriented $Gd_3Ga_5O_{12}$ (GGG) substrates using off-axis sputtering [1,2]. The YIG thin films exhibit pure phase and high crystalline quality as determined by high-resolution x-ray diffraction. As graphene conforms to underlying substrates, it is essential to have smooth YIG films. The typical roughness of grown YIG films is 0.15 nm as confirmed by atomic force microscopy (AFM) [2]. To prepare clean heterostructures of graphene/YIG we employ a dry transfer technique [3,4], where the graphene/YIG interface is never exposed to polymers during the transfer and device fabrication processes. For this, commercially available bulk crystals of hexagonal boron nitride (h-BN) are mechanically exfoliated onto a 90 nm $SiO_2$/Si substrate to get thin flakes of h-BN. The thicknesses of the flakes are confirmed by optical and AFM measurements and a clean 0.6 nm thick h-BN flake is selected for making the heterostructure. This h-BN flake is used to pick-up and transfer graphene and also serves as the tunnel barrier for spin injection [4]. On a separate $SiO_2$/Si substrate, using a residue-free tape, kish graphite is exfoliated to obtain a long and narrow bilayer graphene flake. The bilayer nature of the optically selected flake is further confirmed by Raman spectroscopy [section 2]. A thin polymer, polycarbonate (PC), is coated on a polydimethylsiloxane (PDMS) stamp. Using home built transfer tool, the PDMS stamp with PC is brought in contact with the selected h-BN flake and PC film is softened by slowly heating it to 45 °C. Upon cooling down the PC to room temperature, the h-BN flake gets transferred onto the PC polymer. Then this h-BN flake is carefully aligned and brought in contact with bilayer graphene flake. Due to the strong van der Waals interaction between h-BN and graphene, h-BN picks up the graphene from the substrate. The PDMS/PC stamp with h-BN/graphene is then stamped on a freshly grown YIG surface. By heating to 150 °C, the PC polymer, carrying the h-BN/graphene, is melted onto the YIG substrate which transfers the graphene to YIG. Finally, the PC polymer is dissolved in chloroform. To get rid of any organic/polymer residues on h-BN, the stack is annealed in ultra-high vacuum. The electrodes for electrical spin injection are defined using e-beam lithography and 80 nm Cobalt (Co) is grown in a MBE chamber. The graphene channel length and width between electrodes E2 and E3 are 2.1 μm and 2.2 μm, respectively. The interfacial contact resistances across the 0.6 nm h-BN tunnel barriers are 26 and 34 kΩ for the injector (E2) and detector (E3) respectively.

In order to ensure that the device fabrication process doesn't degrade the magnetic properties of the YIG films, we characterize the YIG substrate after transferring graphene and h-BN. We have used a SQUID magnetometer to measure the magnetic hysteresis of the YIG films. Figure S1 shows an in-plane magnetic hysteresis loop of a YIG film measured at room temperature, which indicates that only a small magnetic field of a few Gauss is needed to saturate the magnetization in the YIG plane.

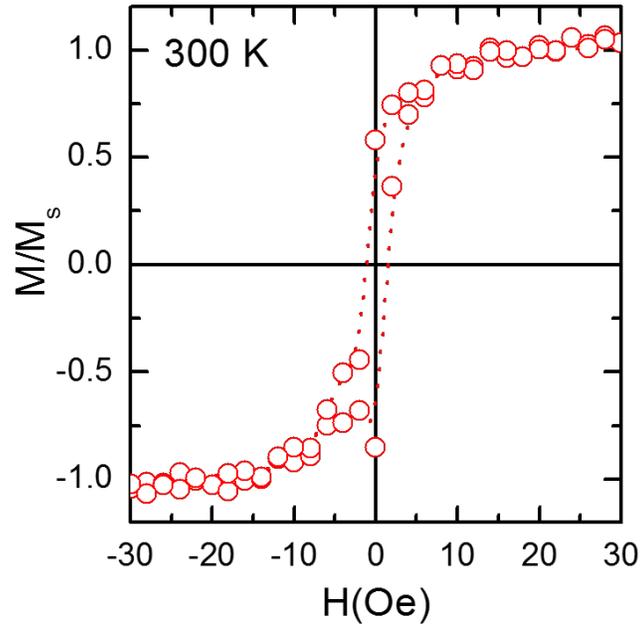

*Figure S1: The magnetic hysteresis loop of YIG film substrate measured at room temperature.*

## 2. Characterization of bilayer graphene

The bilayer nature of graphene is checked by standard optical contrast of an exfoliated flake on a Si substrate with a 300 nm $SiO_2$ layer. To further confirm it, we used Raman spectroscopy on the flake after completing the mechanical transfer of graphene onto the YIG surface. The collected Raman signal excited by a low power, 532 nm green laser is shown in Figure S2. A Raman spectrum over a broad range of wavenumbers is shown in Figure S1a. The Raman peaks of YIG are consistent with previously reports. To confirm the bilayer graphene, more accumulations are recorded near the 2D peak and the measured data is fitted with four Lorentzian peaks, which is a well-established way to confirm bilayer graphene [5,6].

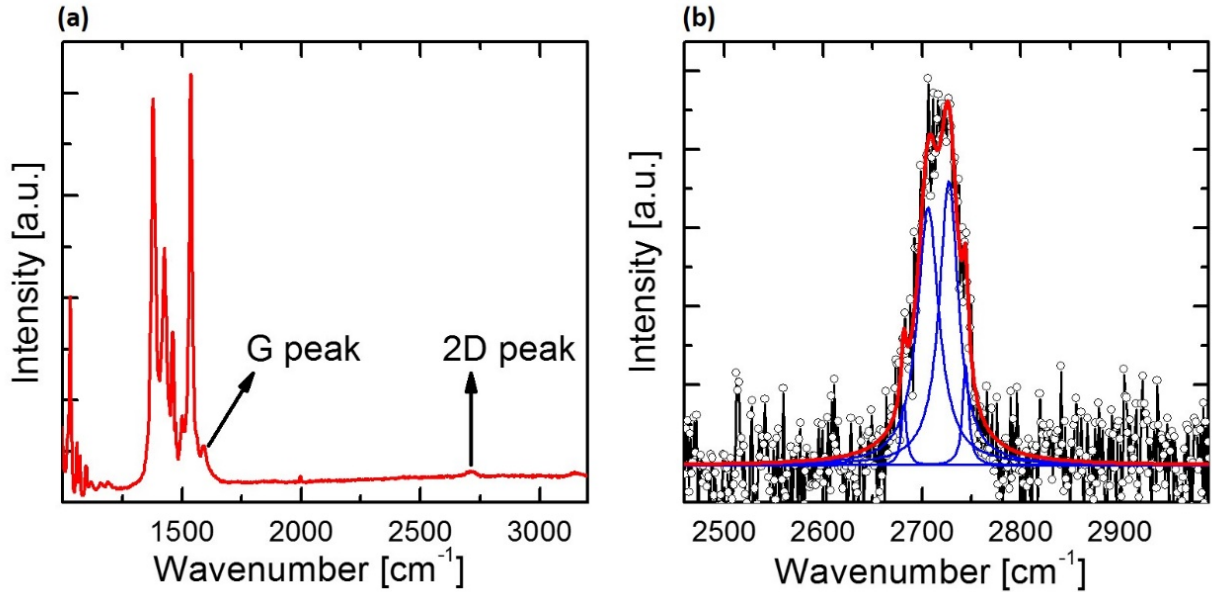

*Figure S2: (a) Raman spectrum collected over a broad wavenumber range with the G and 2D peaks of graphene labelled. (b) Large accumulation Raman spectrum recorded around the 2D peak of graphene. The blue curves shows the four-peak Lorentzian fitting.*

### 3. Control experiment and Anisotropic magnetoresistance of Co electrodes

*3.1 Control experiment with an insulating spacer between graphene and YIG*

Here we provide further evidence that the complete spin modulation observed in graphene/YIG heterostructure is due to the proximity of YIG to the graphene. For this we have prepared a graphene spin valve device which is separated from YIG film by a thin insulating material. A thin insulating spacer in between YIG and graphene should be enough to suppress the magnetic proximity effect, and spin transport in graphene should not be affected by the YIG magnetization. To prepare this heterostructure, we exfoliate a 10 nm thick flake of h-BN (Figure S3b) on a $SiO_2$ substrate. This h-BN flake is transferred onto the YIG substrate following the procedure define in section 1 of the supplementary file. Separately a bilayer graphene flake is exfoliated on a $SiO_2$ substrate (Figure S3a) and then mechanically transferred onto the already prepared h-BN/YIG surface. The graphene/h-BN/YIG heterostructure is shown in Figure S3c. The completed graphene spin valve is shown in Figure S3d, where graphene (on h-BN flake) is highlighted by red dashed lines.

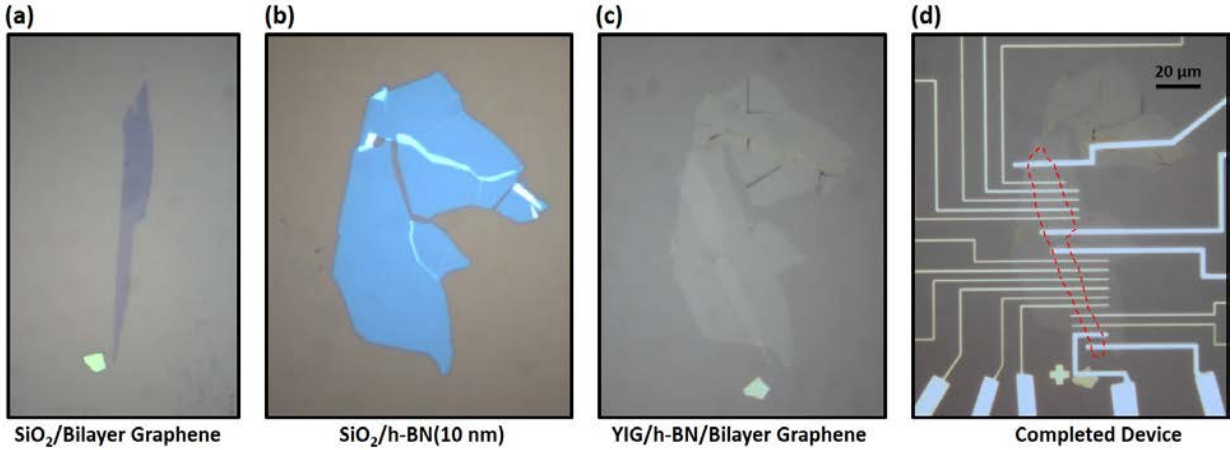

*Figure S3: (a) Exfoliated bilayer graphene on SiO$_2$. (b) Exfoliated h-BN flake on SiO$_2$ (c) Bilayer graphene transferred onto YIG substrate with a 10 nm thick h-BN spacer. (d) The non-local spin valve device with multiple ferromagnetic electrodes. Graphene is highlighted by red dashed lines.*

For the spin modulation experiments, we first measure the non-local MR signal in graphene as explained in the main text. The measured non-local MR is shown in Figure S4a, where a clear signal due to the spin transport in graphene can be seen. To carry out the similar spin modulation experiments as defined in the main text, we apply a small magnetic field in the plane of graphene device and change the angle of the field (θ) while recording the non-local signal. These measurement is carried out for both parallel and anti-parallel alignment of the injector and detector electrodes.

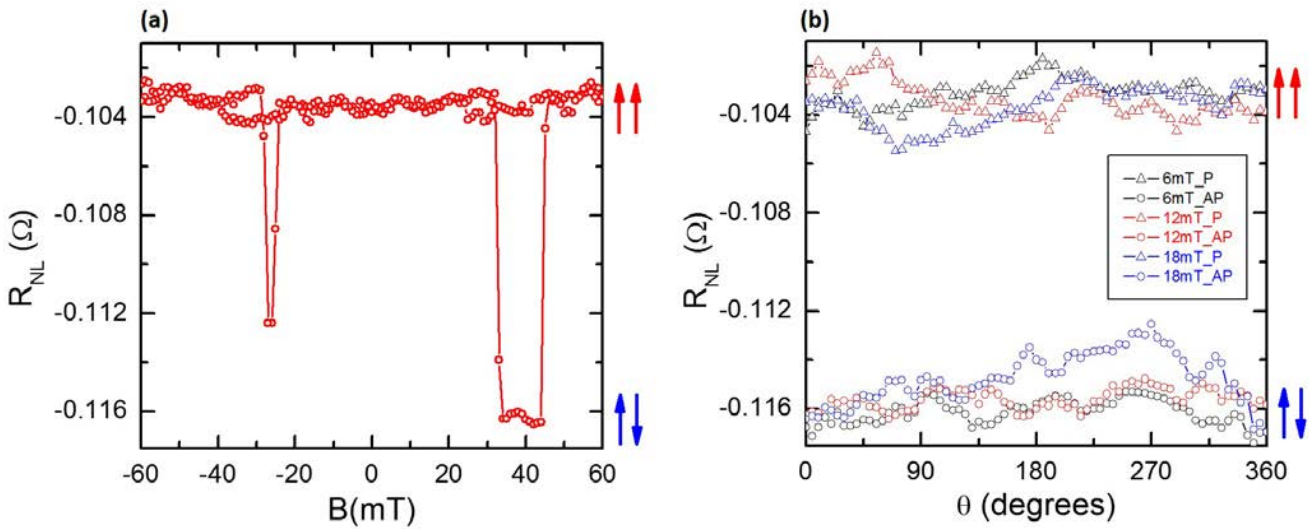

*Figure S4: MR and spin modulation for the controlled device with an h-BN spacer. (a) The measured non-local MR signal in a graphene spin valve, with a h-BN spacer between graphene and YIG. The blue and red arrows represent the relative magnetization direction of injector and detector electrodes. (b) Non-local MR signal measured as function of direction (θ) of applied $B_{ROT}$. The data is shown for different magnitudes of $B_{ROT}$ fields (ranging from 6mT to 18mT). The circles and triangles show the measured data for parallel and anti-parallel configuration of the injector/detector electrodes, respectively.*

In Figure S4b, we plot the observed non-local signal as a function of θ, for $B_{ROT}$ field magnitude varying from 6 mT to 18 mT. As expected, we don't observe the full modulation of the spin signal. For fields as high as 18mT, we observe only a small modulation (within the signal to noise ratio). This observation is consistent with our claim that when graphene is directly coupled to the YIG substrate, the proximity exchange field due to YIG magnetization is responsible for the observed complete spin modulation (for data in the main text). But the addition of a thin insulating layer (in this case h-BN) between graphene and YIG suppresses the proximity and consequently the exchange field experienced by the spins in graphene vanishes.

We also look at the temperature dependence of the MR signals. The MR signal for this control sample is measured in the temperature range ~25K to 300 K and is plotted in the Figure S5. The insets to Figure S5 show the MR signal measured at 25 K and 250 K, with magnitude of MR signal defined as before. As one can notice that there is very weak temperature dependence for the case when graphene is decoupled from the YIG by a thin h-BN layer. This is in stark contrast to the case when graphene is directly placed on YIG. This again is consistent with our claim that the strong temperature dependence of spin signals for the graphene/YIG heterostructure has its origin in the fluctuating exchange fields of YIG magnetization.

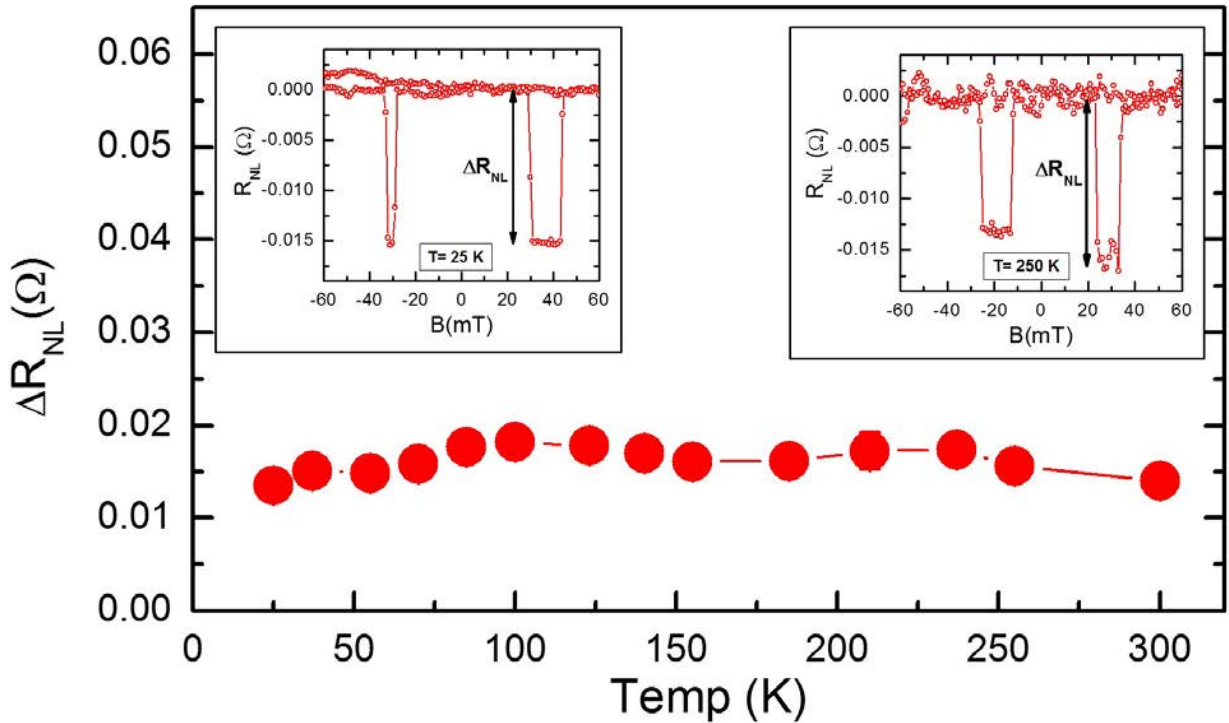

*Figure S5: (a) Temperature dependence of non-local MR signal in a graphene spin valve which is separated from YIG by 10 nm h-BN spacer. The insets to Figure shows the MR signal measured at 25 K and 250 K, where magnitude of the MR is defined by black arrows.*

## 3.2 Anisotropic magnetoresistance of Co electrodes

To further rule out the possibility that the observed 100% spin signal modulation is due to the change in the magnetization direction of Co injector and detector electrodes under the influence of $B_{ROT}$ field, we have performed anisotropic magnetoresistance (AMR) measurements on a Co electrode with dimensions similar to the device injector/detector electrodes. A test Co electrode is electrically connected from opposite ends and a charge current is applied across the Co electrode. Two-probe resistance is measured as a function of the direction of an externally applied magnetic field (in the plane of the electrode) with respect to the current flow direction ($\varphi$) (see the inset to Figure S6). Figure S6 shows the observed magnetic field direction dependent AMR signal. For an applied field of 15 mT, we do not observe any clear change in the Co electrode resistance as the direction of magnetic field is varied. When the field is increased to 150 mT, we see a clear change in the resistance as a function of the applied field direction, which is consistent with the AMR effect. This implies that we don't have considerable change in the magnetization direction of the Co electrode for an applied field of 15 mT in any direction. We expect similar AMR response for the injector and detector Co electrodes of the device used in the main text and thus rule out its contribution to observed spin signal modulation. The observed 100% modulation can only be explained if the magnetization directions of injector and detector electrodes are normal to each other for an applied $B_{ROT}$ field of 15 mT, which is not possible based on Figure S6.

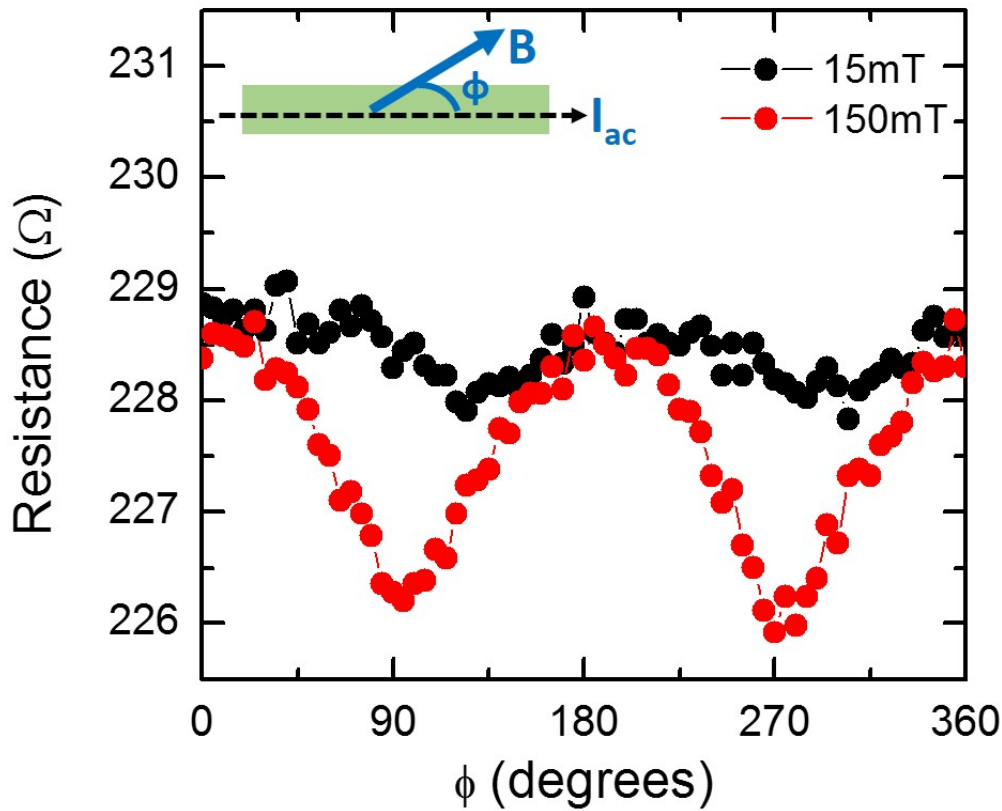

*Figure S6: AMR response of a Co electrode at 15 K as a function of the angle between charge current flow and applied magnetic field (defined in the inset) at magnetic fields of 15 and 150 mT.*

# 4. Modeling of temperature dependent non-local resistance

Spin signal in graphene non-local spin valve (NLSV) on SiO$_2$ normally have a weak temperature dependence. The NLSV signal decrease by factor of 2 (or so) going from 10 K to 300 K [7-9]. However, for graphene on a thin YIG film, we observe that the non-local spin signal rapidly decays as temperature increases, and it disappears at ~230 K. The observed temperature dependence of the spin signal is unusually strong, compared to the weak temperature dependence typically observed for graphene spin valves on non-magnetic substrates [8,10] and require a new physical phenomena to explain our data. In this section, we argue that the observed temperature dependence of the MR spin signal in graphene/YIG can be explained by the electron spin dephasing in graphene due to the random transverse magnetization fluctuations of the YIG. Here we present a detailed discussion of our model to explain the observed temperature dependence data. We start by looking at graphene electron spins dephasing under the influence of static and fluctuating YIG magnetization. In *section 4.1*, we derive a relationship between the spin relaxation rates due to fluctuating magnetization of YIG and write that the transverse component of fluctuating fields cause additional spin dephasing. The spin relaxation rates are expressed in terms of temperature dependent YIG magnetization in the applied field direction ($\bar{M}_y(T)$) and magnetic correlation time ($\tau_c(T)$). As shown in *section 4.2*, $\bar{M}_y(T)$) is obtained from the temperature magnetization characterizations. *Section 4.3* is devoted to obtain $\tau_c(T)$ as function of temperature dependent YIG magnetization ($M_{YIG}(T)$), by adapting a macroscopic picture of local magnetization fluctuation which has been developed through fluctuation-dissipation theorem and had successfully explained spin Seebeck effect in Pt/YIG structure. Next, in *section 4.4* we write the non-local spin signals in graphene by taking into account the additional temperature dependent spin relaxation mechanism due to the presence of fluctuating YIG magnetization. The non-local spin signals are expressed in term of temperature dependent $\bar{M}_y(T)$ and temperature independent parameters: diffusion coefficient of graphene channel ($D$), spin diffusion length ($\lambda_{int}$), spin polarization at Co/h-BN/graphene interface ($p_1p_2$) and damping parameter of YIG. The observed temperature data is fitted and the exchange field as function of correlation time is obtained to approximate a lower bound of exchange field experienced by spins in graphene at graphene/YIG interface. For fittings, a typical value of $D$ is assumed for graphene as we cannot get this from experiments due to lack of electrical gate and Hall-bar configuration of our device. Although $D$, $\lambda_{int}$ and $p_1p_2$ are not absolutely temperature independent parameters [7,10], the small variations of these parameters with temperature cannot alone account for the observed decay of the MR signals in graphene on YIG. Essentially, to limit the number of free parameters in the fit of the temperature dependence, we assume that parameters with weak temperature dependence are constant as a function of temperature. Even under such a constraint, we obtain a very good fit with the experimental data (as shown later in this section), which supports the hypothesis that the strong temperature dependence of the spin signal is due to the spin relaxation induced by a fluctuating exchange field.

To understand the unusual temperature dependent non-local resistance in graphene NLSV on YIG (Fig. 1(b)), we consider the interaction between conduction electron and magnetization of YIG. The terms in the Hamiltonian that associate with conduction electron spin are given by:

$$H_e = A_{ex} \cdot g_e \mu_B \vec{S}_e \cdot \langle \vec{M} \rangle + g_e \mu_B \vec{S}_e \cdot \vec{B}_{app} = g_e \mu_B \vec{S}_e \cdot (\langle \vec{B}_{ex} \rangle + \vec{B}_{app}) = g_e \mu_B \vec{S}_e \cdot \langle \vec{B}_{eff} \rangle \quad (1)$$

where an effective exchange field between conduction electrons in graphene and YIG is defined by

$$\langle \vec{B}_{ex} \rangle = A_{ex} \langle \vec{M} \rangle \quad (2)$$

Here $\vec{M}$ is the YIG magnetization, $A_{ex}$ is the proximity induced exchange coupling strength between YIG and graphene. The averaging $\langle ... \rangle$ is over the ensemble of magnetic moment in YIG that in proximity with graphene.

At finite temperature, $\vec{M}$ in YIG will fluctuate, which leads to fluctuation of exchange field in graphene. In the local picture associate with a single conduction electron travels through graphene, the time and spatial variation of magnetization in YIG will result in a varying effective magnetic field acting on the electron. This varying effective magnetic field can be modelled as a time-dependent, randomly fluctuating magnetic field $\vec{B}_{ex}(t) = \langle \vec{B}_{ex} \rangle + \Delta \vec{B}_{ex}$. Previous theoretical work had predicted that the randomly fluctuating magnetic field can cause additional spin relaxation [11,12], and the model had successfully explained spin transport phenomena in graphene with adatom carrying paramagnetic magnetic moment [13]. Furthermore, the fluctuation strength of magnetization in YIG is temperature dependent. As a result, the spin relaxation rate caused by the magnetization fluctuation can also be temperature dependent. In the following, we are going to use the above model to understand our temperature dependence data.

### 4.1 Spin relaxation induced by YIG magnetization fluctuation

For the non-local geometry (inset to Figure 4a, main text), the injected spin polarization, the applied magnetic field, and the effective exchange field lie along the same axis (y axis in our case). The spin relaxation rate induced by the random fluctuating field is given by the longitudinal spin relaxation term:

$$\frac{1}{\tau_1^{ex}} = \frac{(\Delta B_{tr})^2}{\tau_c} \frac{1}{(B_{app,y} + \bar{B}_{ex,y})^2 + (\gamma_e \tau_c)^{-2}} \approx \frac{(\Delta B_{tr})^2}{\tau_c} \frac{1}{(\bar{B}_{ex,y})^2 + (\gamma_e \tau_c)^{-2}} \qquad (3)$$

where $(\Delta B_{tr})^2 = (\Delta B_{ex,x})^2 + (\Delta B_{ex,z})^2$ is the fluctuation of exchange field in the transverse direction. $B_{app,y}$ is ignored as $\bar{B}_{ex,y} \gg B_{app,y}$. $\gamma_e$ is the gyromagnetic ratio of electron. $\tau_c$ is the correlation time of exchange field fluctuations defined as:

$$\langle \Delta \vec{B}_{ex}(t) \cdot \Delta \vec{B}_{ex}(t - t') \rangle_t \propto exp(-t/\tau_c) \qquad (4)$$

The exchange field fluctuation in graphene is strongly associated with the magnetization fluctuation of YIG through (2). At finite temperature, thermal fluctuation suppresses the equilibrium magnetization from 0 K value, and induce a fluctuating magnetization. On a statistical basis, the equilibrium magnetization and fluctuating magnetization follows a simple rule:

$$\langle M_0^2 \rangle = \langle M_x^2 + M_y^2 + M_z^2 \rangle = (\bar{M}_y)^2 + (\Delta M_x)^2 + (\Delta M_y)^2 + (\Delta M_z)^2 \qquad (5)$$

where $M_0$ is the saturation magnetization of YIG at 0 K, and the equilibrium magnetization is in the y direction. Assuming the transverse magnetization fluctuation in YIG dominates, we have:

$$(\Delta M_x)^2, (\Delta M_z)^2 \gg (\Delta M_y)^2 \qquad (6)$$

As noted in (2), the effective exchange field in graphene is proportional to magnetization in YIG. We can rewrite the exchange field and transverse exchange field fluctuation from (5) (6) as:

$$(\Delta B_{tr})^2 = (A_{ex}\Delta M_{tr})^2 = A_{ex}^2((\Delta M_x)^2 + (\Delta M_z)^2) = A_{ex}^2\left((M_0)^2 - \left(\bar{M}_y\right)^2\right) \tag{7}$$

$$\bar{B}_{ex,y} = A_{ex}\bar{M}_y \tag{8}$$

Equation (3) can be rewrite as:

$$\frac{1}{\tau_1^{ex}} = \frac{A_{ex}^2\left((M_0)^2 - \left(\bar{M}_y(T)\right)^2\right)}{\tau_c\left[\left(A_{ex}\bar{M}_y(T)\right)^2 + (\gamma_e \tau_c(T))^{-2}\right]} \tag{9}$$

where both $\bar{M}_y(T)$ and $\tau_c(T)$ depends on temperature.

### 4.2 Temperature Dependence of $\bar{M}_y$

We can extract the temperature dependence of $\bar{M}_y$ from experiment. We measured temperature dependence of saturation magnetization of YIG with various techniques up to 300 K (Fig. 2(a)). Previous study of bulk YIG shows that the reduction of saturation magnetization follows $\sim T^{3/2}$ in the low temperature regime (<25 K), while follows a $\sim T^3$ in the higher temperature regime (25 K~250 K) [14]. We fit with data with both term, and find that the contribution of $T^3$ term is negligible (Figure 4d main text). To simplify the spin transport equation later, we simply assume that:

$$\frac{\bar{M}_y}{M_0} = 1 - aT^{\frac{3}{2}} \tag{10}$$

and get $a = 6.314\times10^{-5} K^{-3/2}$ from fitting with experimental data.

### 4.3 Temperature Dependence of $\tau_c$

A macroscopic picture of local magnetization fluctuation had been developed by through fluctuation-dissipation theorem[15] and had successfully explained spin Seebeck effect in Pt/YIG structure [16-18]. In the following, we are going to adapt the model into our system to approximate the correlation time. Consider an ensemble of localized magnetic moment M at the graphene/YIG interface. The dynamics of this moment can be modeled by the Landau-Lifshitz-Gilbert (LLG) equation:

$$\partial_t\vec{M} = \gamma(\vec{H}_0 + \vec{h})\times\vec{M} + \frac{\alpha}{M_s}\vec{M}\times\partial_t\vec{M} \tag{11}$$

Where $\vec{H}_0 = H_0\hat{z}$ is the effective magnetic field on the localized moment, which is proportional to magnetization of the YIG film ($M_{YIG}$). $\gamma$ is the gyromagnetic ratio, $\alpha$ is the Gilbert damping constant, and $M_s$ is the saturation magnetization. We further define $\vec{h}$ as the noise field. By the fluctuation-dissipation theorem [19,20], $\vec{h}$ is assumed to obey the following Gaussian ensemble [15]:

$$\langle h^\mu(t)\rangle = 0 \tag{12}$$

$$\langle h^\mu(t) h^\nu(t') \rangle = \frac{2k_B T_F \alpha}{\gamma a_S^3 M_s} \delta_{\mu,\nu} \delta(t-t') \tag{13}$$

Where $a_s^3 = \hbar\gamma/M_s$ is the cell volume of the ferromagnet.

In order to calculate the dynamics of $\vec{M}$, we consider the transverse component of (11):

$$\partial_t M_x = -\gamma(H_{0z} + h_z)M_y + \gamma h_y M_z + \frac{\alpha}{M_s} M_y \partial_t M_z - \frac{\alpha}{M_s} M_z \partial_t M_y \tag{14}$$

$$\partial_t M_y = \gamma(H_{0z} + h_z)M_x - \gamma h_x M_z - \frac{\alpha}{M_s} M_x \partial_t M_z + \frac{\alpha}{M_s} M_z \partial_t M_x \tag{15}$$

Consider (14) and (15) only up to the linear term, by defining $m^\mu = M^\mu/M_s$ and assume that $H_0 \gg h_z$, we obtain:

$$\partial_t m_x = -\gamma H_0 m_y - \alpha \partial_t m_y + \gamma h_y \tag{16}$$

$$\partial_t m_y = \gamma H_0 m_x + \alpha \partial_t m_x - \gamma h_x \tag{17}$$

The heart of this calculation is to study the dynamics of the fluctuation, which is represented by the quantity:

$$\langle \Delta\vec{m}(t) \cdot \Delta\vec{m}(t') \rangle = \langle m_x(t)m_x(t') + m_y(t)m_y(t') \rangle = \frac{1}{2}(\langle m^+(t)m^-(t') \rangle + \langle m^-(t)m^+(t') \rangle) \tag{18}$$

Utilizing the fact that $\langle \Delta\vec{m}(t) \cdot \Delta\vec{m}(t') \rangle$ is only a function of $t - t'$, we can rewrite (18) as:

$$\langle \Delta\vec{m}(t) \cdot \Delta\vec{m}(t') \rangle = \int_{-\infty}^{\infty} \frac{d\omega}{4\pi} \langle m_\omega^+ m_{-\omega}^- + m_\omega^- m_{-\omega}^+ \rangle \tag{19}$$

To solve $m_\omega^+$ and $m_\omega^-$, we can rewrite (16),(17) as:

$$\partial_t(m_x + im_y) = i\gamma H_0(m_x + im_y) + i\alpha\partial_t(m_x + im_y) - i\gamma(h_x + ih_y) \tag{20}$$

$$\partial_t(m_x - im_y) = -i\gamma H_0(m_x - im_y) - i\alpha\partial_t(m_x - im_y) + i\gamma(h_x - ih_y) \tag{21}$$

This leads to that

$$\partial_t m^+ = i\gamma H_0 m^+ + i\alpha\partial_t m^+ - i\gamma h^+ \tag{22}$$

$$\partial_t m^- = -i\gamma H_0 m^- - i\alpha\partial_t m^- + i\gamma h^- \tag{23}$$

With Fourier transform that $f(t) = \int \frac{d\omega}{2\pi} f_\omega e^{-i\omega t}$, and define $\omega_0 = \gamma H_0$, we can rewrite the above two equations in the Fourier representation:

$$\omega m_\omega^+ = -\omega_0 m_\omega^+ + i\alpha\omega m_\omega^+ + \gamma h_\omega^+ \tag{24}$$

$$\omega m_\omega^- = \omega_0 m_\omega^- - i\alpha\omega m_\omega^- - \gamma h_\omega^- \tag{25}$$

We can obtain that:

$$m_\omega^+ = \frac{\gamma h_\omega^+}{\omega_0 + \omega - i\alpha\omega} \tag{26}$$

$$m_\omega^- = \frac{\gamma h_\omega^-}{\omega_0 - \omega - i\alpha\omega} \tag{27}$$

Insert (26) and (27) back to (19), we obtain:

$$\langle \Delta \vec{m}(t) \cdot \Delta \vec{m}(t') \rangle = \int_{-\infty}^{\infty} \frac{d\omega}{4\pi} \langle \frac{\gamma^2 h_\omega^+ h_{-\omega}^-}{(\omega+\omega_0)^2+\alpha^2\omega^2} + \frac{\gamma^2 h_\omega^- h_{-\omega}^+}{(\omega-\omega_0)^2+\alpha^2\omega^2} \rangle e^{i\omega(t-t')} \quad (28)$$

From (13), we can obtain that:

$$\int_{-\infty}^{\infty} \frac{d\omega}{2\pi} \langle h_\omega^+ h_{-\omega}^- \rangle e^{-i\omega(t-t')} = \int_{-\infty}^{\infty} \frac{d\omega}{2\pi} \langle h_\omega^- h_{-\omega}^+ \rangle e^{-i\omega(t-t')} = \frac{4 k_B T_F \alpha}{\gamma a_S^3 M_s} \delta(t-t') \quad (29)$$

We can obtain that:

$$\langle h_\omega^+ h_{-\omega}^- \rangle = \langle h_\omega^- h_{-\omega}^+ \rangle = \frac{8\pi k_B T_F \alpha}{\gamma a_S^3 M_s} \quad (30)$$

Inserting (30) back into (28), we can obtain that:

$$\langle \Delta \vec{m}(t) \cdot \Delta \vec{m}(t') \rangle = \frac{4\gamma k_B T_F \alpha}{a_S^3 M_s} \cos\left(\frac{\omega_0}{1+\alpha^2}(t-t')\right) e^{-\frac{\alpha}{\sqrt{1+\alpha^2}}\omega_0(t-t')} \quad (31)$$

The correlation time can be obtained as:

$$\frac{1}{\tau_c} = \frac{\alpha}{\sqrt{1+\alpha^2}} \omega_0 = \frac{\alpha \gamma H_0}{\sqrt{1+\alpha^2}} = \eta M_{YIG}(T) \quad (32)$$

where $\eta$ is a proportionality constant.

### 4.4 Simplify the Expression of $\tau_1^{ex}$

To simplify the expression of $\tau_1^{ex}$, we can first put (32) and (10) back to (9):

$$\frac{1}{\tau_1^{ex}} = \frac{A_{ex}^2\left((M_0)^2-(\bar{M}_y(T))^2\right)}{\tau_c\left[(A_{ex}\bar{M}_y(T))^2+(\gamma_e \tau_c(T))^{-2}\right]} = \frac{1-(\bar{M}_y/M_0)^2}{\bar{M}_y/M_0} \cdot \frac{\eta(\gamma_e A_{ex})}{(\gamma_e A_{ex})^2+\eta^2} \cdot \gamma_e A_{ex} M_0 \quad (33)$$

Notice that only the first term depends on temperature. We can define the first term as $\xi(T)$, and rewrite the whole equation as:

$$\frac{1}{\tau_1^{ex}} = \xi(T) \cdot \frac{\eta(\gamma_e A_{ex})}{(\gamma_e A_{ex})^2+\eta^2} \cdot \gamma_e A_{ex} M_0 \quad (34)$$

### 4.5 Implement $\tau_{ex}$ into non-local spin transport

Consider non-local spin transport in graphene spin valve with tunneling contact [7,21]:

$$R_{NL} = p_1 p_2 R_N e^{-L/\lambda} \quad (35)$$

where $\lambda = \sqrt{D\tau_{total}}$ is the spin diffusion length in graphene, D is the diffusion coefficient, $R_N$ is the spin resistance of the graphene channel and $\tau_{total}$ is the spin life time. The spin dephasing rate can be written as:

$$\frac{1}{\tau_{total}} = \frac{1}{\tau_1^{ex}} + \frac{1}{\tau_{other}} \quad (36)$$

where $\frac{1}{\tau_1^{ex}}$ is the spin relaxation caused by magnetization fluctuation, and $\frac{1}{\tau_{other}}$ is the spin relaxation caused by other mechanism. $\frac{1}{\tau_{other}}$ can be considered as the spin relaxation of graphene on a non-magnetic substrate. The spin diffusion length in graphene can be rewritten as:

$$\lambda = \sqrt{D\tau_{total}} = \left(\frac{1}{D\tau_{total}}\right)^{-\frac{1}{2}} = \left(\frac{1}{D\tau_1^{ex}} + \frac{1}{D\tau_{other}}\right)^{-\frac{1}{2}} = \left(\frac{1}{D\tau_1^{ex}} + \frac{1}{\lambda_i^2}\right)^{-\frac{1}{2}} \quad (37)$$

where $\lambda_i$ is the spin diffusion length of graphene on a non-magnetic substrate. The non-local spin transport of graphene on YIG is then given as:

$$R_{NL} = p_1 p_2 R_N e^{-L \cdot \left(\frac{1}{D\tau_1^{ex}} + \frac{1}{\lambda_{int}^2}\right)^{-1/2}} \quad (38)$$

and

$$\frac{1}{D\tau_1^{ex}} = \frac{\gamma_e}{D}\xi(T) \cdot \frac{\eta(\gamma_e A_{ex})}{(\gamma_e A_{ex})^2 + \eta^2} \cdot A_{ex}M_0 = \frac{\xi(T)}{\beta} \quad (39)$$

by defining

$$\frac{1}{\beta} = \frac{\gamma_e}{D}\frac{\eta(\gamma_e A_{ex})}{(\gamma_e A_{ex})^2 + \eta^2} \cdot A_{ex}M_0 \quad (40)$$

The non-local spin transport equation can be written as:

$$R_{NL} = \mathcal{R}\, e^{-L \cdot \left(\frac{\xi(T)}{\beta} + \frac{1}{\lambda_{int}^2}\right)^{-1/2}} \quad (41)$$

We fit the temperature dependence of the non-local spin signal (Figure 4a of the main text) using equation 41 with $\mathcal{R}, \lambda_{int}, \beta$ as the three fitting parameters, and taking $L = 2.1\ \mu m$ and $\xi(T)$ determined from the temperature dependence of the YIG magnetization (Figure 4d of the main text). The model fits very well with the experimental data and yields best fit parameters: $\mathcal{R} = 0.7015\ \Omega, \lambda_{int} = 1.9561\ \mu m, \beta = 1.5578 \times 10^{-13}\ m^2$. The fittings with different parameters: Channel length (L), intrinsic spin diffusion length ($\lambda_{int}$) and $\beta$ are shown in Figure S7, S8 and S9 respectively.

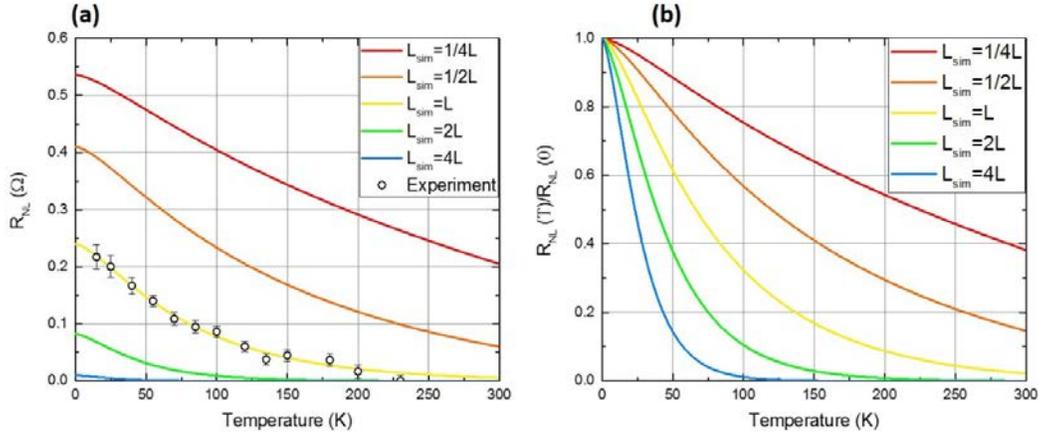

Figure S7: (a) Simulation of temperature dependent non-local signal with different graphene channel length. (b) Same data normalized with 0 K value.

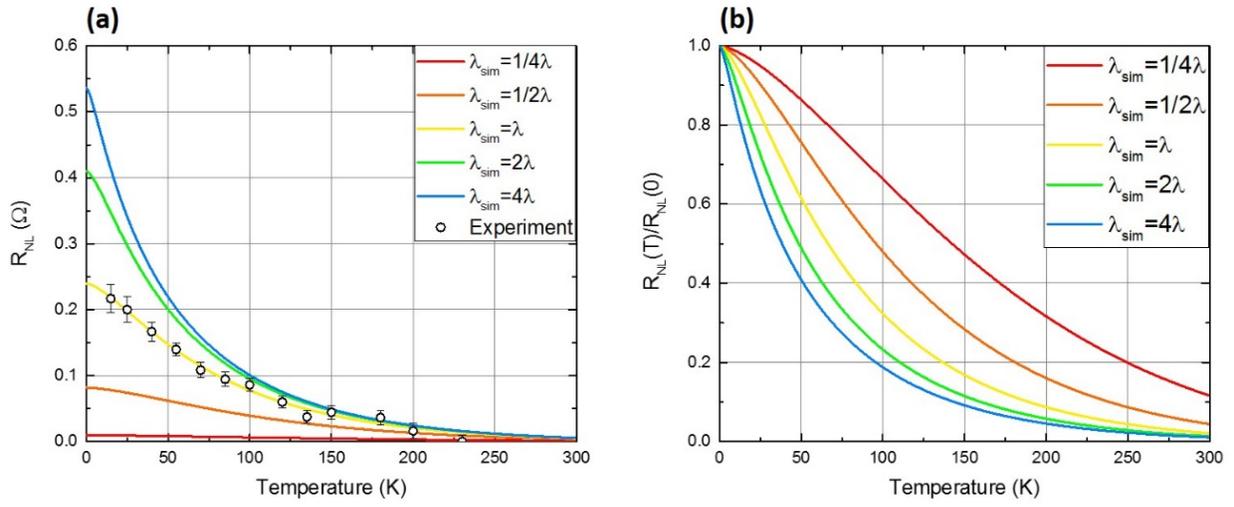

*Figure S8: (a) Simulation of temperature dependent non-local signal with different graphene intrinsic spin diffusion length. (b) Same data normalized with 0 K value.*

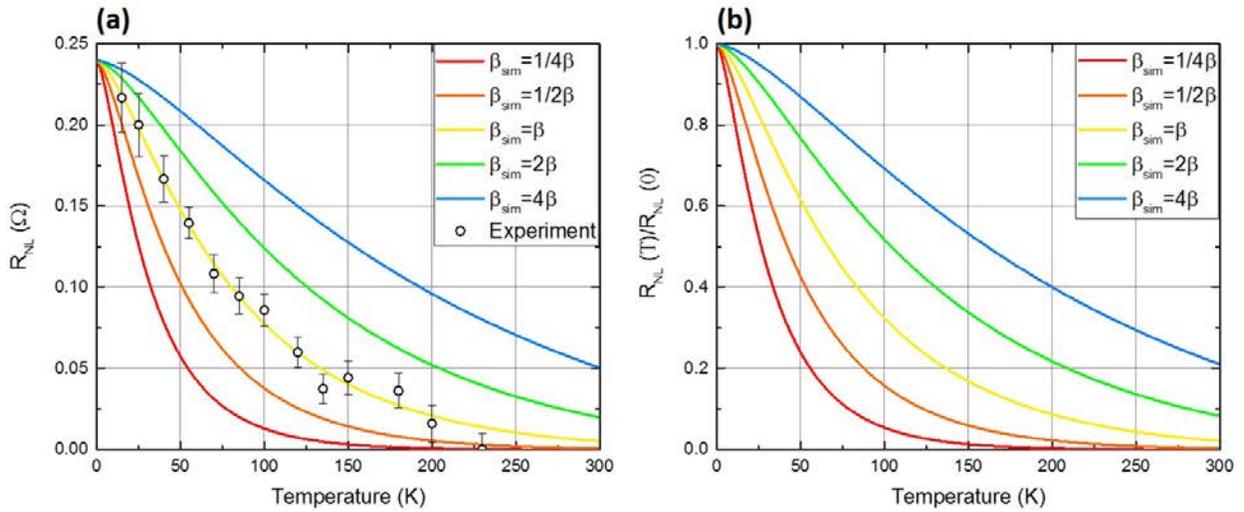

*Figure S9: (a) Simulation of temperature dependent non-local signal with different value of the β coefficient in the fitting. (b) Same data normalized with 0 K value.*

To calculate the exchange field in graphene at 0 K, we look at the $\beta$ coefficient from the fitting. We can rewrite (40) as:

$$\eta \beta M_0 (\gamma_e A_{ex})^2 - D(\gamma_e A_{ex})^2 = D\eta^2 \tag{42}$$

which leads to

$$B_{ex}(0) = A_{ex} M_0 = \sqrt{\frac{D}{\eta \beta M_0 - D}} \cdot \frac{\eta M_0}{\gamma_e} = \frac{1}{\gamma_e} \cdot \sqrt{\frac{1}{\left(\frac{\beta}{D}\right)\tau_c - \tau_c^2}} \tag{43}$$

where $\tau_c$ is the correlation time at 0 K. Assuming $D = 0.015\ m^2/s$, we can plot $B_{ex}(0)$ as function of different $\tau_c$ (Fig. 4(e) main text).

## 5. Hanle measurements

In this section, we provide the Hanle measurements for graphene spin valve on YIG substrate as well as for graphene on non-magnetic substrates. For the case of graphene on YIG, the in-plane Hanle measurements are performed by applying a transverse magnetic field along the x-axis (see Figure 1d, main text). The inset to Figure S10a shows the schematic of the in-plane Hanle measurement setup. The non-local signal is measured as function of applied field for both parallel and anti-parallel configuration of the ferromagnetic electrodes. The difference of parallel and anti-parallel signal is plotted in Figure S10a. In fact, the in-plane Hanle is essentially the same as the spin modulation experiments shown in the main text. For the out of plane Hanle measurements, a magnetic field is applied perpendicular to the plane of the graphene device and the non-local signal is measured as function of applied field. The measured signal is plotted in the Figure S10b.

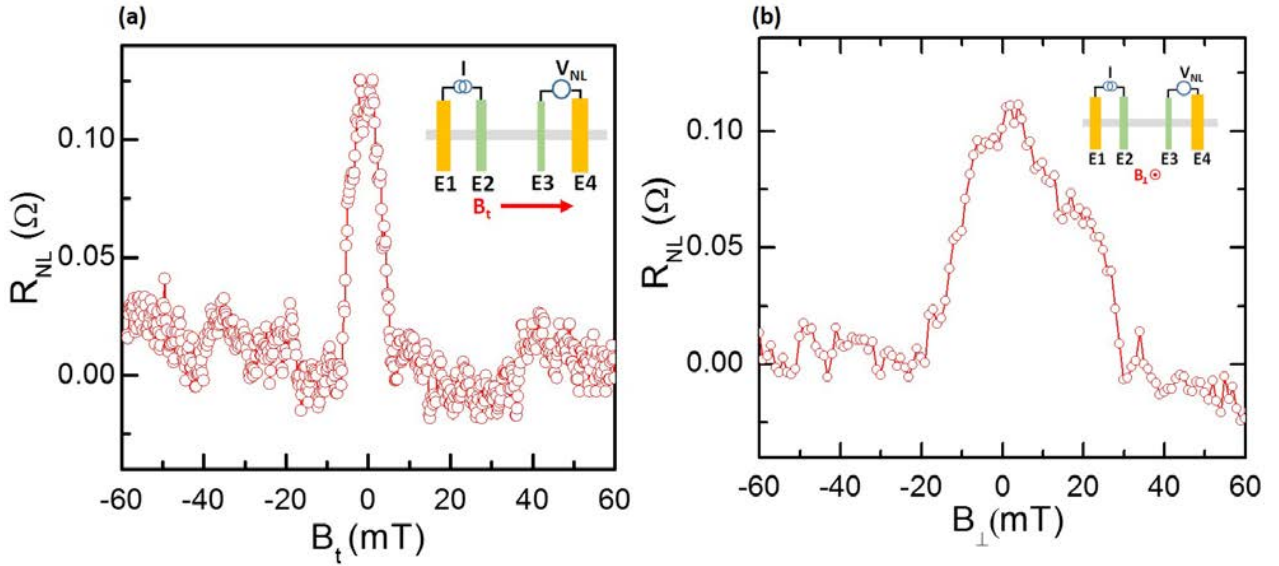

*Figure S10: Hanle measurements for graphene on YIG (a) In-plane Hanle curve measured by applying a magnetic field transverse to the length of Co electrodes. (b) Out of plane Hanle curve, showing the non-local resistance as a function of the magnetic field applied normal to the graphene plane. Inset to the Figures shows the schematic of the Hanle measurements.*

Also, as the spin relaxation in graphene does not have large anisotropy (i.e. whether spins are precessing in graphene plane or out of plane) [22], the wider out of plane Hanle (as compared to in-plane Hanle) in graphene/YIG heterostructures can be qualitatively understood from the picture of proximity exchange field as explained below: During the in-plane Hanle measurements, very small applied fields (few mT) are sufficient to fully saturate YIG magnetization and hence the spins are under the influence of large net field (sum of applied and magnetic exchange) resulting in a sharp in-plane Hanle curve. For out of plane Hanle, the YIG magnetization is not fully saturated (it requires ~200 mT to saturate out of plane as confirmed by out of plane magnetization measurements) and thus only a small component of maximum proximity field is acting on graphene spins making the out of plane Hanle curves wider than the corresponding in-plane Hanle curves.

Next, we present the Hanle precession measurements for control experiments i.e. graphene on non-magnetic substrates. Figure S11a and Figure S11b show the in-plane and out of plane Hanle measurements, respectively, for the graphene spin valve on $SiO_2$ control sample used in Figure 3b and 3c of the main text. Note that unlike graphene on YIG, the width and shape of the Hanle curves are similar for both in-plane and out of plane spin precession measurements. This is consistent with previous experimental reports of isotropic spin relaxation in graphene on non-magnetic substrates [22].

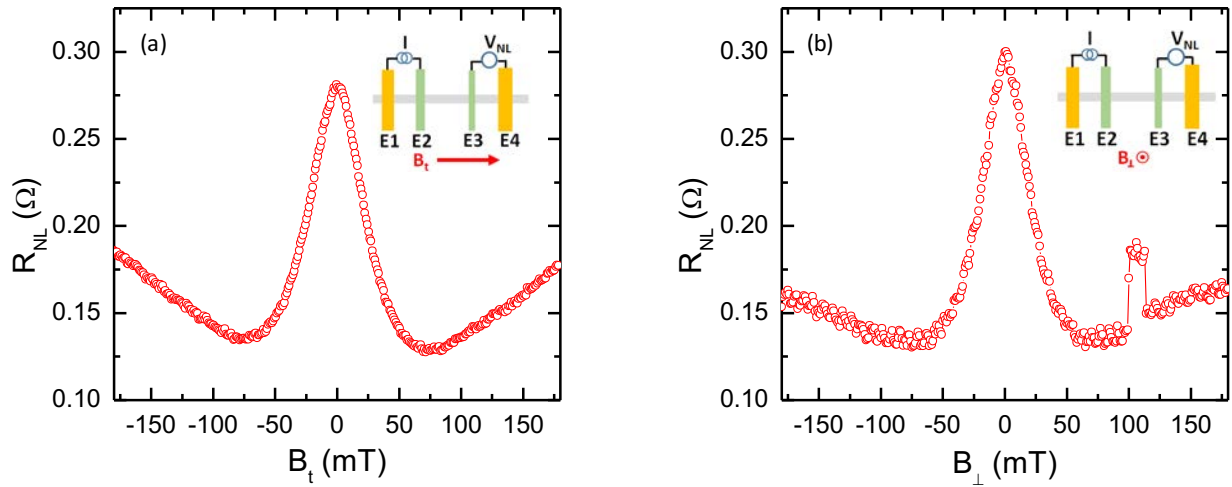

*Figure S11: Hanle measurements for graphene on $SiO_2$ substrate. (a) In-plane Hanle curve measured by applying a field perpendicular to the injected spins. (b) Out of plane Hanle curve measured by applying a magnetic field normal to graphene. The jump at ~100 mT is likely due to a Co magnetization change in a high perpendicular field. Inset to Figures (a) and (b) depicts the measurement setup. Data shown are for parallel alignment of injector and detector magnetizations.*

Also, it is important to notice that for the case of graphene on YIG, there is very sharp Hanle curve (Figure S10) in comparison to the very smooth Hanle curves for graphene on non-magnetic substrates (Figure S11). For graphene on YIG, the net field around which spins in graphene precess is the sum of the externally applied magnetic field and the magnetic exchange field originating from the proximity of YIG. As explained earlier this makes the in-plane Hanle curve for graphene on YIG super sharp as compared to in-plane Hanle precession curves for graphene on a non-magnetic substrate (like $SiO_2$). Thus, qualitatively, the sharper decay of the in-plane and out of plane Hanle curves for graphene on YIG provides additional experimental evidence of the magnetic exchange field in graphene/YIG heterostructures.

We have also carried out the Hanle precession measurements in a graphene channel which is separated from YIG by a 10 nm thick insulating h-BN layer. Figure S12 shows the in-plane Hanle curve which looks similar to the Hanle curve for graphene spin valves on $SiO_2$ substrates. As expected, the shape and width of the Hanle curve looks like what one would expect for graphene on a non-magnetic substrate [10,22]. This observation again suggests that by inserting an insulating material (h-BN) between YIG and graphene, the magnetic proximity effect can be quenched completely.

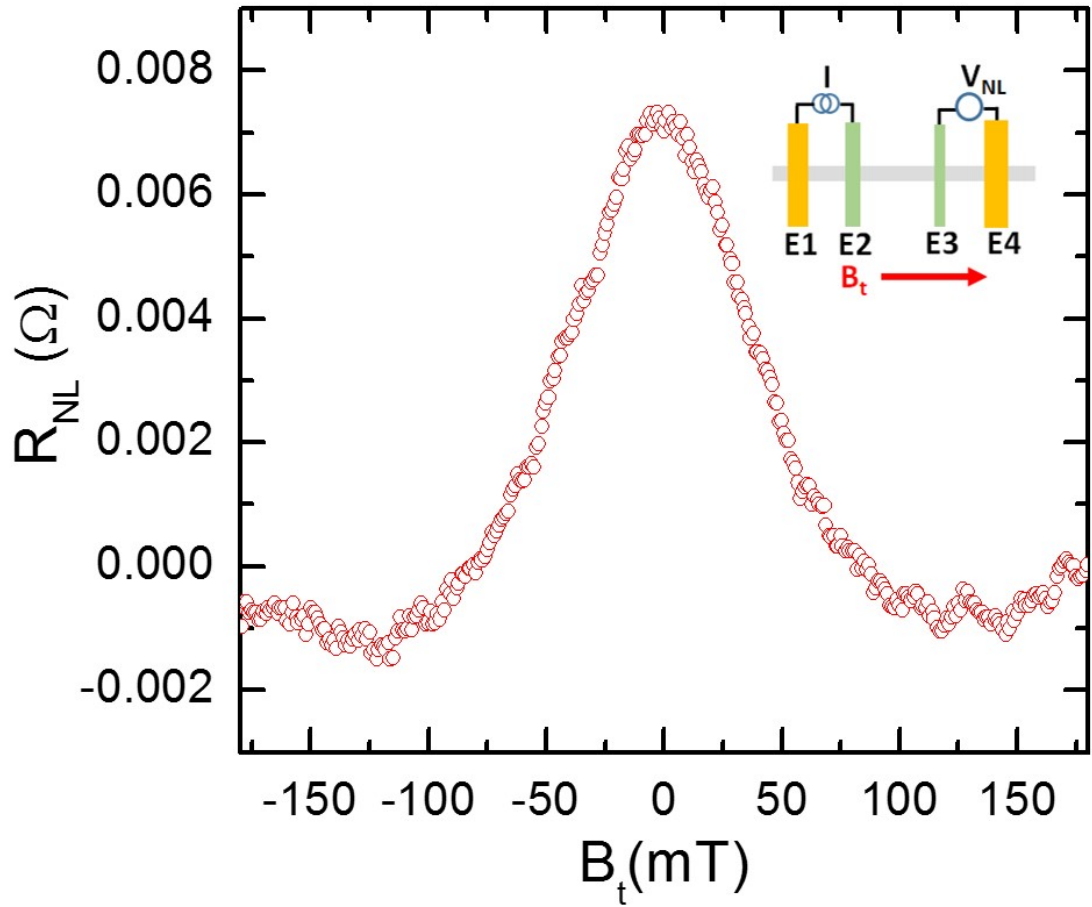

*Figure S12: In-plane Hanle measurements for graphene channel separated from YIG substrate by 10 nm thick insulating h-BN spacer. Inset to Figure shows the schematic of Hanle measurement.*

# 6. Temperature dependence of the spin signal modulation

We also present the temperature dependence of the spin signal modulation measured up to 150 K. Above 150 K, the non-local MR signals are vanishingly small and the due to poor signal to noise ratio we couldn't resolve any modulation data. Figure S8, shows the non-local MR signal and corresponding spin signal modulation for both parallel and anti-parallel configurations. As it is evident, for all the measured temperatures we see a complete spin signal modulation at all temperatures irrespective of the fact that the spin signal is dying off fast due to increasing spin dephasing owing to the transverse magnetization fluctuations of YIG. This points to the effectiveness and persistence of interfacial magnetic proximity effect at temperatures as high as 150 K.

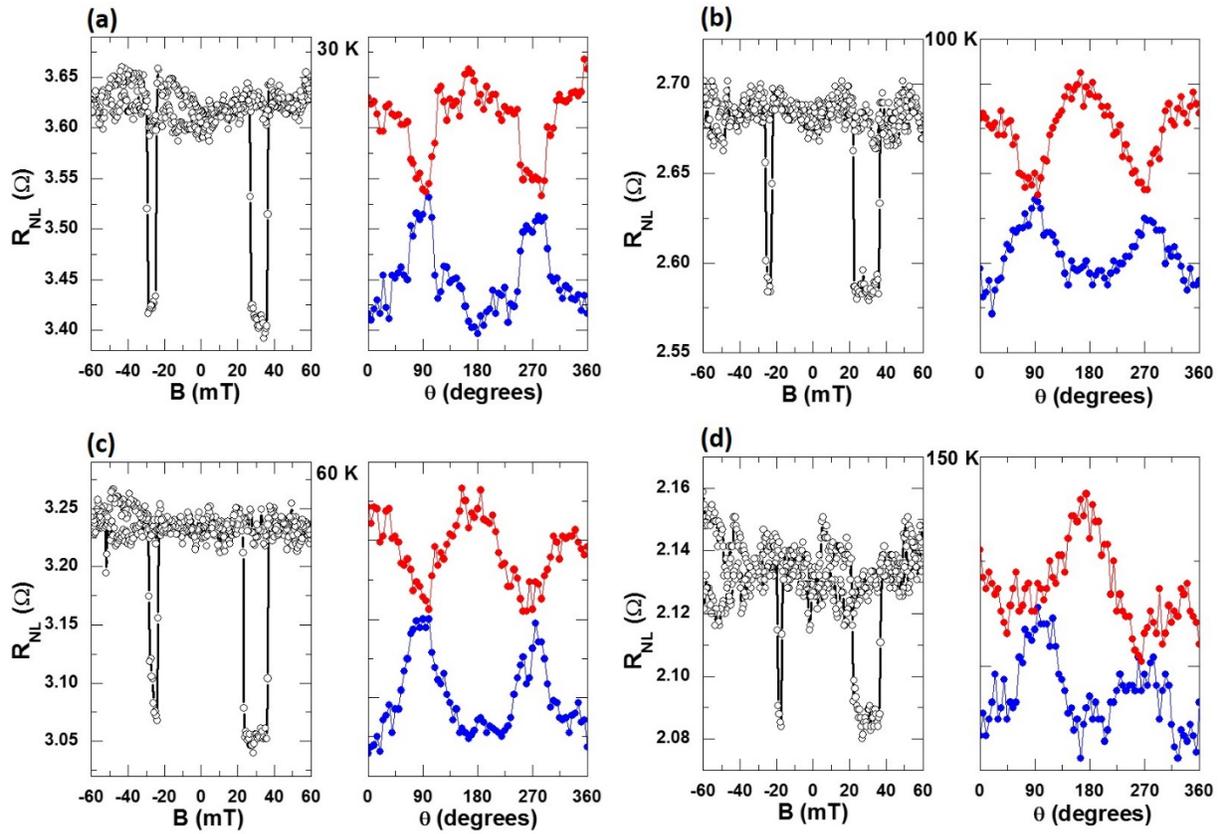

*Figure S7: (a), (b), (c) and (d) shows the temperature dependence of the measured MR and spin modulation signal at different temperatures. The sample temperature is denoted in each figure. The red and blue data points in each figure shows data for parallel and anti-parallel configuration of injector and detector electrodes respectively.*